\UseRawInputEncoding
\documentclass[journal,a4paper]{IEEEtran}
\usepackage{graphicx} 
\usepackage{amsmath}
\usepackage{float}
\usepackage{subfig}
\usepackage{caption}
\usepackage{url}
\usepackage{hyperref}

\usepackage[boxed,ruled,vlined,linesnumbered,commentsnumbered, noresetcount]{algorithm2e}
\usepackage{algorithmic}

\usepackage{diagbox}
\usepackage{tikz}
\usetikzlibrary{shapes.geometric}
\usetikzlibrary{positioning}

\usepackage{soul}

\usepackage{tabularx}
\usepackage{ragged2e} 
\newcolumntype{L}{>{\RaggedRight}X} 
\usepackage{lipsum} 
\usepackage{amsmath}
\usepackage{amssymb}
\usepackage{comment}
\usepackage{makecell} 
\usepackage{multirow}

\usepackage{pifont}
\newcommand{\xmark}{\ding{55}}%

\makeatletter
\newcommand\notsotiny{\@setfontsize\notsotiny\@vipt\@viipt}
\makeatother

\begin{document}

\title{A Comprehensive Survey of Wireless Time-Sensitive Networking (TSN): Architecture, Technologies, Applications, and Open Issues}

\author{Kouros Zanbouri, Md. Noor-A-Rahim, Jobish John, Cormac J. Sreenan,~\IEEEmembership{Senior Member, IEEE,} H.~Vincent~Poor,~\IEEEmembership{Fellow, IEEE,} and~Dirk Pesch,~\IEEEmembership{Senior Member, IEEE}

\thanks{Kouros Zanbouri, Md. Noor-A-Rahim,  Cormac Sreenan, and Dirk Pesch are with the  School of Computer Science \& IT, University College Cork,  Ireland.  (E-mail: {\tt \{k.zanbouri, m.rahim, cjs, d.pesch\}@cs.ucc.ie}). 

Jobish John is with the  Eindhoven University of Technology, Netherlands. 

H. V. Poor is with the  Department of Electrical and Computer Engineering, Princeton University, USA (E-mail: {\tt poor@princeton.edu}).

This publication has emanated from research conducted with the financial support of Science Foundation Ireland under Grant number 13/RC/2077\_P2. The work of Vincent Poor was supported by the U.S National Science Foundation under Grant ECCS-2335876.

Corresponding Author: Md Noor-A-Rahim

}}

\maketitle

\begin{abstract}
Time-sensitive networking (TSN) is expected to be a key component of critical machine-type communication networks in areas such as Industry 4.0, robotics and autonomous vehicles. With rising mobility requirements in industrial applications and the prevalence of wireless networks, wireless network integration into TSN is becoming increasingly important. This survey article presents a comprehensive review of the current literature on wireless TSN, including an overview of the architecture of a wireless TSN network and an examination of the various wireless technologies and protocols that can be or are used in such networks. In addition, the article discusses industrial applications of wireless TSN, among them industrial automation, robotics, and autonomous vehicles.  The article concludes by summarizing the challenges and open issues related to the integration of TSN into wireless networks, and by offering suggestions for future research directions.

\end{abstract}



\section{Introduction}


Time-critical communication has been a key feature of various communication networks such as in industrial automation systems \cite{8714025, 10034532}, in real-time applications for healthcare \cite{10035961}, autonomous driving \cite{8743176}, professional audio/video delivery in an Internet of Multimedia Things \cite{s20082334}, smart-grids \cite{FAHEEM2021106854}, and other applications, where highly reliable and timely data delivery are required to ensure high Quality of Service (QoS). In such networks, wireless communication links and device mobility can result in a significant degree of unreliability \cite{8839862,9815179}. 

IEEE~802.1 Time-Sensitive Networking (TSN) is a collection of standards and tools to facilitate real-time deterministic communications with low latency and high reliability \cite{9881899,9483592}. While TSN aims to be physical and data link layer agnostic, current implementations are essentially all based on Ethernet as the network medium. The market for time-sensitive networking, which was valued at USD 0.2 billion in 2023, is anticipated to grow to USD 1.7 billion by 2028, at a Compound Annual Growth Rate (CAGR) of 58.3\% \cite{tsnmarket2023}. The use of TSN ensures minimal congestion loss for time-sensitive data traffic, thanks to resource reservations, multiple queuing mechanisms, ultra-high reliability measures, and traffic shaping methods. TSN guarantees that important data will not exceed a maximum end-to-end latency, enabling dependable data transfer by implementing reliability mechanisms at the packet level. Additionally, TSN safeguards against threats such as bandwidth misuse, equipment failure, and malicious cyberattacks, among others \cite{8412457}. 

By extending the features of TSN to wireless communication, wireless TSN enables more flexible and scalable communication systems that can be used in locations where wired networks are not feasible or practical, resulting in reduced installation and maintenance expenses \cite{9652097}. Thus, wireless TSN aims at delivering the same low-latency and high-reliability features that are available with wired TSN, while accommodating applications that require wireless communications. In the context of enabling TSN for wireless networks, several TSN standards have been developed to address issues such as the need for precise time synchronization over IEEE~802.11 networks, as specified in the IEEE~802.1AS standard \cite{8290700}. Although the majority of the approaches in the TSN standards are not specifically tailored to data link layer constraints, this broad specification allows for future adaptation and modification to accommodate widely used wireless network protocols, such as IEEE~802.11, 5G, as well as future 6G. This paves the way for wireless TSN to be employed in various domains, including industrial automation, autonomous driving, healthcare, and multimedia applications. 
However, in contrast to traditional wired communications, wireless communication introduces a range of challenges such as erroneous, non-reciprocal channels, signal distortion, latency, and radio interference (both self-interference and external interference). This makes wireless communication unpredictable and complicates the assurance of performance requirements and bounds as usually guaranteed by wired TSN. Hence, by their nature, wireless networks are not ideal to support real-time critical traffic and extremely low-latency applications. 

In this paper, we provide a comprehensive survey of the existing literature on wireless TSN, with a focused exploration of its architecture, technologies, standards, applications, and pertinent challenges in its integration with wireless networks. Previous, related surveys have mostly delved into generic TSN, leaving a notable gap in the exploration of wireless TSN. The authors of \cite{fedullo2022comprehensive,john2023industry} provide a thorough analysis of the TSN Working Group's standards and protocols whilst investigating industrial TSN use cases, concentrating especially on distributed measurement systems in industry. The last section of the paper is dedicated to TSN usage in Wi-Fi networks, as Wi-Fi networks are in widespread use in industrial environments. Adame at. al. \cite{adame2021time} discuss the essential characteristics and components of IEEE~802.11be (Wi-Fi 7) and demonstrate how to utilize TSN functions with Wi-Fi 7. The paper also presents the main applications and use cases of Wi-Fi 7 for IoT with low latency. This work covers only Wi-Fi 7, while IEEE~802.11ac and IEEE~802.11ax, which play a significant role in current wireless network deployments, are not covered, representing an important omission. The authors in \cite{seol2021timely} reviewed recent TSN research, including the most recent IEEE standards. The paper classified articles based on their subject, objective, and technique to analyze the current state of TSN-related research. The paper highlights directions for overcoming the limitations of TSN for real-world deployments. In \cite{8458130}, a comprehensive survey explores IEEE TSN and IETF DetNet, categorizing them by flow concepts, synchronization, management, control, and integrity. It also discusses achieving ultra-low latency in 5G networks within three key categories: front haul, backhaul, and network management, while highlighting current standards and research limitations. Atiq et. al. \cite{9652097} analyzed the IEEE~802.11 and 5G standards and integration with wired TSN with the goal of providing low-latency, predictable communications through this integration. Moreover, the authors highlight use cases such as automation, transportation, and wireless TSN for audio/video applications, underscoring the limited attention given to wireless TSN technologies in existing surveys.  The authors in \cite{SATKA2023102852}, provided a concise overview of current research on integrating TSN with 5G network. The paper details the methodology of their systematic review, including planning, execution, and analysis. This analysis further reveals emerging trends, technical requirements, and areas where existing research could be bolstered, thereby charting a course for future investigations in this domain. The existing survey works are summarized in the following Table \ref{Table:relatedwork}:

\begin{table}[htbp]
\caption{Comparison of existing wireless TSN surveys}  \label{Table:relatedwork}    
\scriptsize
\centering          
\begin{tabular}{|c|c|c|}  
\hline 
\textbf{Work} & \textbf{Wireless TSN Coverage} & \textbf{Differentiating aspects} \\[0.5ex]  
\hline \hline
\cite{fedullo2022comprehensive} & TSN over Wi-Fi & - Limited assessment \\ \hline
\cite{seol2021timely} & \makecell{TSN over Wi-Fi \\ TSN over 5G } & \makecell{- Superficial assessment of presented works \\ - Not evaluating technical aspects } \\ \hline
\cite{8458130} & TSN over 5G & - Not evaluating Wi-Fi potentials \\ \hline
\cite{SATKA2023102852} & TSN over 5G & \makecell{ - Only focused on 5G } \\ \hline

\end{tabular}
\end{table}

\normalcolor

To the best of our knowledge, this is the first time a comprehensive survey has been devoted solely to the examination of the literature on wireless TSN. Furthermore, existing surveys do not delve deeply into the integration of TSN with more recent versions of the IEEE~802.11 standard, such as IEEE~802.11be, or with 6G. To bridge this knowledge gap and to encourage further research and development in the field of wireless TSN, this paper presents a comprehensive survey of the current state-of-the-art in wireless TSN.

\begin{table}[htbp]
\caption{List of Acronyms and Definitions}
\begin{tabular}{|l|l|}
\hline
Abbreviation & Description \\
\hline
3GPP & 3rd generation partnership project \\
5G & Fifth Generation Cellular \\
AGV & Automated Guided Vehicles \\
ATS & Synchronous Traffic Shaping \\
AVB& Audio Video Bridging \\					
BLE & Bluetooth Low Energy \\				
CSMA/CA & Carrier Sense with Multiple Access Collision Avoidance  \\
CNC & Centralized Network Configuration \\
CNCD & Computing Network Collaboration Domains \\
CUC & Centralized User Configuration \\
FGA & Fine-grained aggregation \\
FPGA & Field-Programmable Gate Arrays \\
FRER & Frame Replication and Elimination for Redundancy \\
FTM & Fine Timing Measurement \\
GCL & Gate Control List \\
GM & Grand Master \\
gPTP & generic Precision Time Protocol \\				
MAC & Medium Access Control \\	
MCS & Modulation and Coding Scheme \\
MLO & Multi-Link Operation \\
NIST & National Institute of Standards and Technology \\
OFDM & Orthogonal Frequency-Division Multiplexing \\				
OT &  Operating Technology \\
PER & Packet Error Rate \\
PHY & Physical Layer \\
PLC & Programmable Logic Controller \\
QoE & Quality of Experience \\
QoS & Quality of Service \\	
SINR & Signal to Interference and Noise Ratio \\			
TM & Timing Measurement \\
TSN & Time-Sensitive Networking \\
TSNC & Time-Sensitive Networking Controller \\
TTE & Time-Triggered Ethernet \\
UE & User Equipment \\	
UNI & User Network Interface \\	
URLLC & Ultra-Reliable Low-Latency Communications \\
\hline
\end{tabular} \label{tab:Abbr}
\end{table}

\begin{figure*}[t]
\centering
\includegraphics[scale = 0.5]{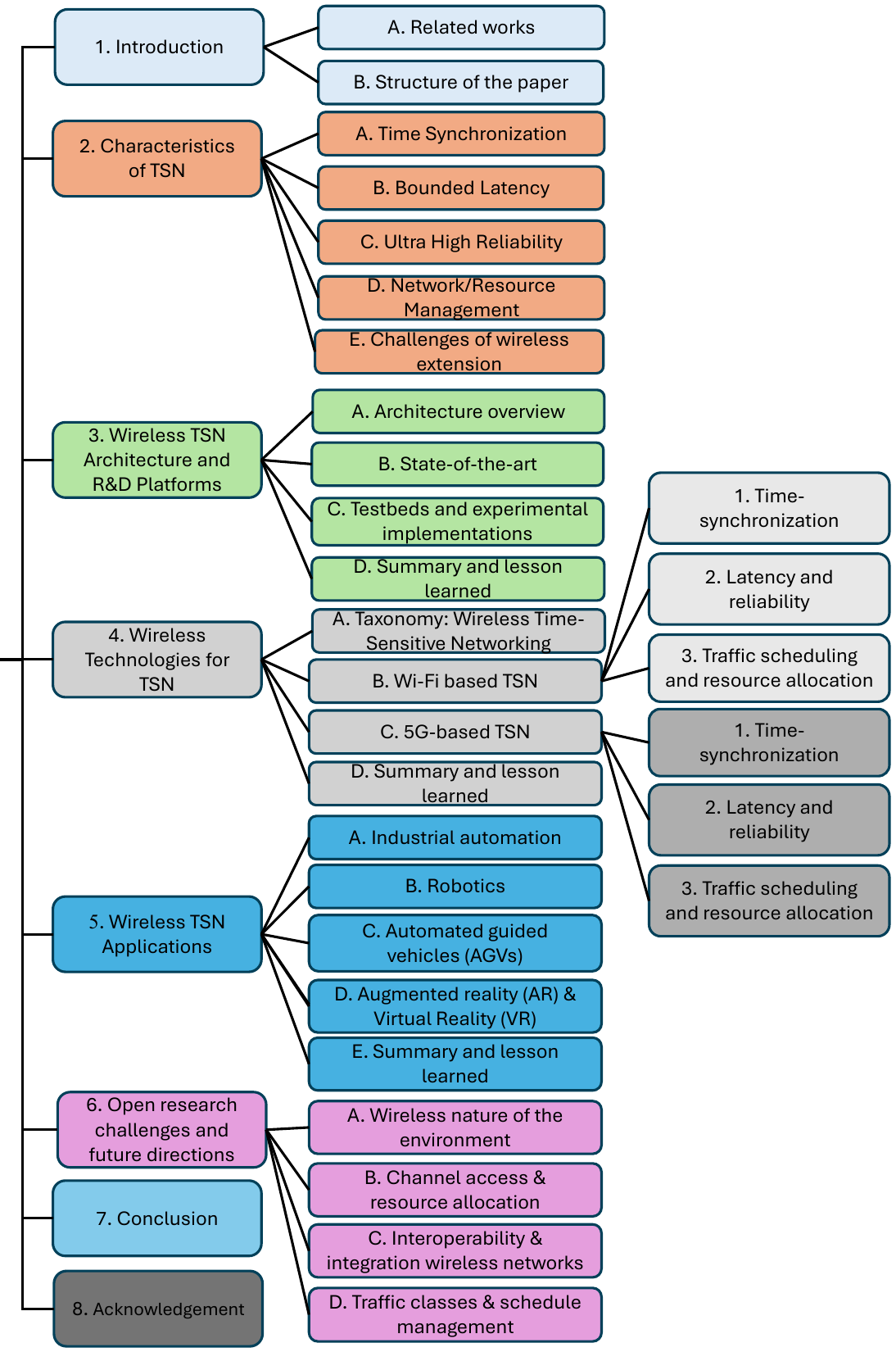}
\caption{Structure of this paper}
\label{fig_PaperArch}
\end{figure*}

The paper is structured as follows: Section~\ref{sec:background} provides an overview of the background pertaining to traditional wired TSN, its underlying characteristics, and industrial network applications, which have motivated the development of wireless TSN. Section~\ref{sec:Arch} delves into the wireless TSN architecture and its fundamental components. In Section~\ref{sec:Tech}, we present an analysis of various wireless technologies that are suitable for TSN integration, highlighting their primary features and implementation methods for enabling TSN capabilities. 
Section~\ref{sec:App} investigates the applications of wireless TSN in various fields and use cases. Section~\ref{sec:Open} then discusses open research challenges and issues, along with possible directions for future research. Finally, we conclude this article in Section~\ref{sec:Con}. The structure of the paper is presented in Fig.~\ref{fig_PaperArch}.

\begin{figure*}[t]
\centering
\includegraphics[scale = 0.5]{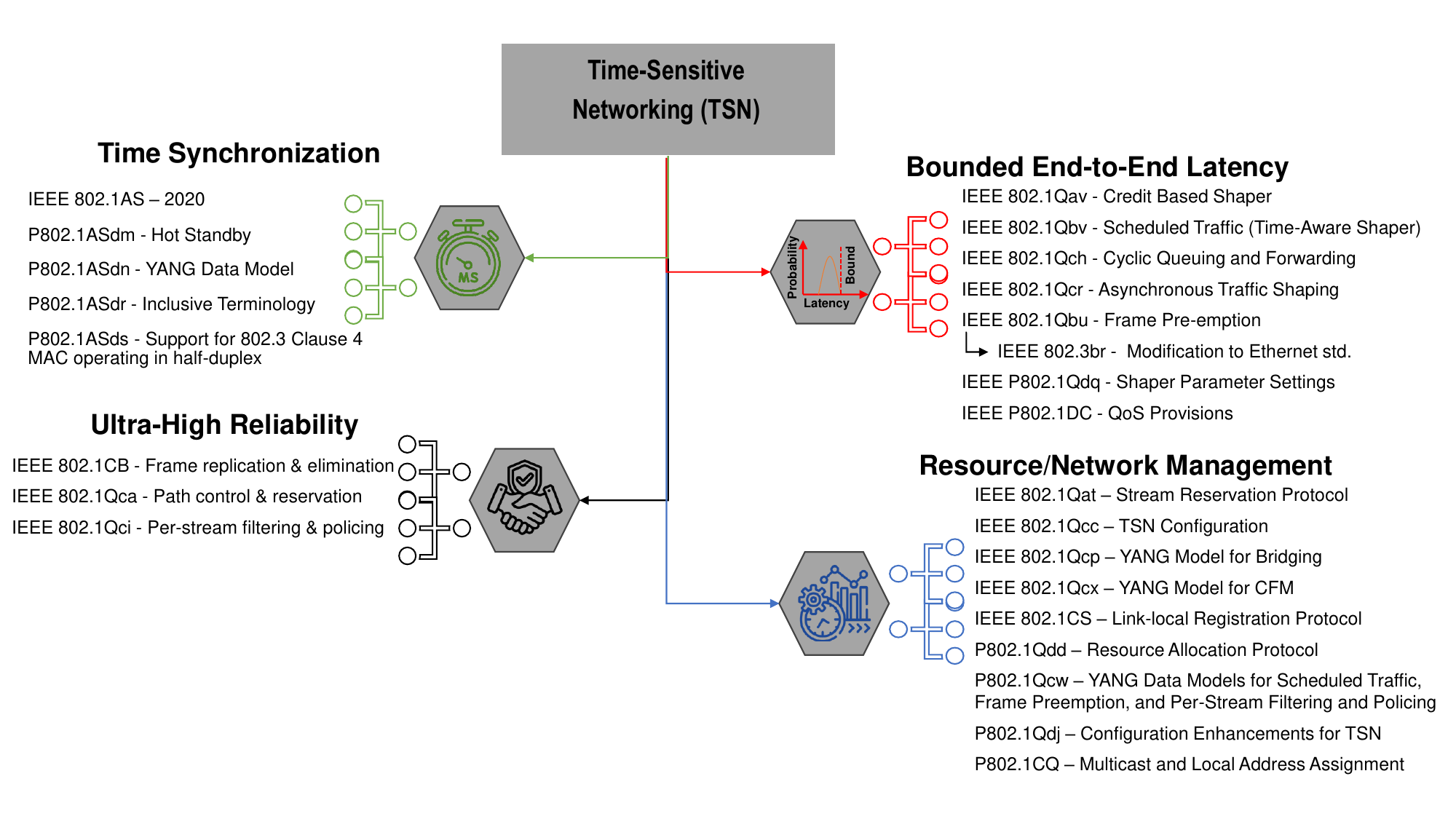}
\caption{TSN; Features, key mechanisms \& associated standards}
\label{fig_tsn}
\end{figure*}

\section{Characteristics of TSN} \label{sec:background}
The IEEE~802.1 TSN Task Group, formed in 2012 \cite{TSNtaskGroup}, has been working to establish Ethernet standards that incorporate improvements for real-time applications with high-reliability demands. TSN is mainly composed of modifications and amendments to the IEEE~802.1Q standard. Due to its widespread industry acceptance and the development of TSN switches by multiple manufacturers, TSN has been gaining traction as a dependable networking standard in various fields such as manufacturing, healthcare, multimedia, robotics, and smart grids. The main applications and characteristics of TSN are depicted in Fig \ref{fig_tsn}. The TSN standards are classified according to their characteristics, which are discussed in detail in the following subsections. TSN relies on the interaction of various entities to ensure reliable and deterministic communication. Talkers are the sources of information generating data packets based on particular application requirements. Listeners, on the other hand, receive these packets and process them by forwarding them to their respective applications. A continuous stream of data packets flowing between a talker and a listener forms a communication channel. Each data unit transmitted on the network is encapsulated within a frame, comprising a header containing metadata and a payload carrying the actual data. This structured approach enables TSN to convey the information for necessary guarantees for time-sensitive applications.


\subsection{Time Synchronization} 

Time synchronization plays a key role in guaranteeing the predictable performance of end devices and their time-sensitive applications. For example, typical control cycles in factory automation range from 100 to 1000 milliseconds with a jitter of less than 1 millisecond. In contrast, robotic control systems can have control cycles up to 1 millisecond with a jitter of less than 1 microsecond \cite{8672474}.

The IEEE~802.1AS standard on timing and synchronization for time-sensitive applications was created as part of a series of standards for Audio/Video Bridging (AVB) by the 802.1 task group \cite{4659212,ieee2020ieee}. IEEE~802.1AS offers a generic precision time protocol (gPTP) profile to achieve precise time synchronization throughout the network using the IEEE 1588v2 Precision Time Protocol (PTP) \cite{mildner2019time,10.1145/3273905.3273920}. To enable strictly bounded end-to-end latency and several other features, TSN requires precise time synchronization up to sub-microsecond levels of accuracy among all network elements such as switches, bridges, and end devices (manufacturing robots, PLCs, etc.). The Precision Time Protocol uses Ethernet frames in a distributed fashion to synchronize the timing of all network elements to that of the leader device (known as Grandmaster-GM). Decent synchronization can be achieved with a software-based PTP implementation that was found to attain a mean synchronization accuracy of 30 microseconds and a standard deviation of 54 microseconds \cite{9169833}.

The time synchronization mechanism is based on a leader-follower (master-slave) approach, and all devices in the network are synchronized to the grandmaster (the device with the most accurate time source in the network) through periodic exchange of sync and follow-up messages over a spanning tree.
 Using the PTP protocol, the master clock broadcasts time synchronization messages (Sync) to which slaves respond with requests for more timing information (Delay-Req). The master then sends additional messages (Follow-Up and Delay-Resp) to help slaves calculate the time difference between their clocks and the master's clock. By averaging the time it takes for messages to travel between the master and slave, slaves estimate the network delay and refine their clock synchronization \cite{10012852}. 
The need for TSN time synchronization is generated by applications that require bounded latency, which influences network scheduling. Usually, a network must maintain strict synchronization to deliver the lowest possible or strictly bounded latency for time-sensitive traffic \cite{9524299}. To transmit time to an end application across for example IEEE~802.11 connections, either the Timing Measurement (TM) or the Fine Timing Measurement (FTM) protocol, which are specified in IEEE~802.11-2016, are utilized \cite{10034503}. The IEEE~802.1AS protocol operation over IEEE~802.11 differs slightly from its operation over Ethernet. In both scenarios, both TM and FTM functionalities facilitate the calculation of time offset and frequency between stations (STAs). This is achieved by estimating neighbour ratios and propagation delays through time stamping of IEEE~802.11 action frames and acknowledgement frames exchanged among STAs \cite{10012852}.

\subsection{Bounded Latency} TSN-based networks provide bounded low latency with zero congestion loss communications to prioritize time-sensitive packets while being transmitted across the network. To achieve this, various TSN mechanisms, such as traffic scheduling, traffic shaping, and frame pre-emption, have been established under the IEEE~802.1 series of standards such as 802.1Qav, 802.1Qbv, 802.1Qbu, 802.1Qch, and 802.1Qcr \cite{8412458}. These mechanisms can handle different traffic classes to achieve different QoS in multi-hop switched networks. IEEE~802.1Qbv manages time-critical traffic flow based on a time-triggered scheduling approach, while IEEE~802.1Qav is meant to handle traffic flows with relaxed latency bounds by adopting a "credit" based transmission strategy to allocate bandwidth \cite{atiq2021ieee}. 
IEEE~802.1Qbv specifies that each traffic class (queue) is assigned a timed gate, capable of being opened or closed based on a predefined Gate Control List (GCL). This timed gate mechanism, known as Time-Aware Shaper (TAS), is implemented at the egress ports of network devices. TAS uses a predetermined timetable, defined in Gate Control Lists (GCLs), to control when data for each traffic type can be sent.

Data is exclusively transmitted from the designated queue only when its corresponding gate is open. In cases where multiple gates are simultaneously open, frames from the highest priority queue are given precedence for transmission, potentially blocking others until either the queue becomes empty or its associated gate closes \cite{9463925}. Furthermore, the standard incorporates a lookahead mechanism for every traffic class, allowing assessment of whether sufficient time remains to transmit the entire frame before the gate's closure. If insufficient time is available, the frame cannot be forwarded until the subsequent opening of the window, resulting in an idle period (guard band) at the conclusion of the current window opening \cite{9916254}.

Additionally, frame pre-emption as outlined in IEEE~802.1Qbu and IEEE~802.3br, allows high-priority frames, known as express frames, to interrupt the transmission of lower-priority frames in Ethernet networks, thereby improving network efficiency and minimizing latency \cite{cavalcanti2020wireless}. In traditional Ethernet networks without frame preemption, a data packet of maximum size (1500 Bytes) can occupy a network port for approximately 12.5 microseconds at a speed of 1 Gigabit per second. However, with frame preemption, this blocking time can be significantly reduced to about 1 microsecond. When a frame is interrupted by a higher-priority frame, the remaining parts of the interrupted frame are referred to as fragments. These fragments must be at least 64 Bytes in size. To prioritize frame handling, network ports maintain separate queues for express (high-priority) and preemptible (lower-priority) frames. These frames are processed by dedicated components known as express MAC and preemptible MAC \cite{9067122}.
A new method for managing network traffic, called Cyclic Queuing and Forwarding (CQF), has been introduced in the IEEE~802.1Qch standard, which combines the features of earlier standards, IEEE~802.1Qbv and IEEE~802.1Qci, to ensure consistent delay between network points by sending and holding time-sensitive data in a regular pattern along the network. CQF is a simpler alternative to TAS. CQF works by sending and holding time-sensitive data in a regular pattern along the network. This approach guarantees predictable delivery times without the complex setup required by TAS \cite{9407828}. As a result, building and managing TSN systems using CQF is easier than using older methods. This makes TSN technology more suitable for situations that demand flexibility and constant change. The IEEE~802.1Qch cyclic queuing and forwarding system offers the advantage of determining packet latency by summing per-hop delays if the cycle time is adjusted appropriately \cite{9125493}. Similarly, IEEE~P802.1Qdq and IEEE~P802.1DC standards provide guidelines for shaper parameters in handling bursty traffic and QoS functionalities within the network, respectively \cite{8412458}.

When it comes to traffic shaping, although TAS is effective in enforcing predictable traffic patterns, the demanding timing specifications, especially the need for precise timing synchronization throughout the TSN network, introduce greater complexity and pose a risk to the network's reliability in the event of any timing discrepancies \cite{8681083}. The Asynchronous Traffic Shaping (ATS) (IEEE~802.1Qcr \cite{9253013}), an alternative to TAS, aims to achieve low congestion loss and deterministic performance within the TSN domain, without relying on time synchronization \cite{zhou2019insight}.

\subsection{Ultra-High Reliability}
Ultra-high reliability is a crucial requirement for TSN-enabled networks to ensure the delivery of each packet without any loss or congestion-related delays. 
Applications that are highly time-sensitive and safety-critical cannot afford the re-transmission of lost packets. To address this, the IEEE~802.1Qca and IEEE~802.1CB standards define redundancy mechanisms for TSN networks. IEEE~802.1CB Frame Replication and Elimination for Reliability, or FRER for short, provides redundancy by sending packets along multiple paths and eliminating duplicates at the receiving end \cite{8569454}.  Every duplicated flow travelling along a separate route is referred to as a member stream. All member streams carrying identical data are categorized as a single compound stream. Within each member stream, duplicate packets are recognized by an Individual Recovery Function (IRF). Duplicates arriving from distinct member streams are distinguished through the Sequence Recovery Function (SRF). The SRF assigns sequence IDs to packets before duplicating them across various member streams. Subsequently, the SRF facilitates the removal of duplicate packets \cite{9838905, 10390539}. 
On the other hand, the IEEE~802.1Qca protocol offers routing functions to handle bridged networks, enabling the establishment of different paths and allocating resources along those paths. This protocol creates multiple routes within the network, providing precise control over these paths to ensure reliable data delivery between devices. Additionally, it reserves network capacity for specific data streams and manages the communication process for efficient and synchronized data transmission \cite{9212162, 9652097}. These parts of the standards series ensure reliability in TSN networks and support time-critical traffic flows without relying on re-transmission mechanisms. Additionally, the IEEE~802.1Qci standard provides per-stream filtering and policing mechanisms to safeguard time-sensitive traffic by prioritizing queued frames. This involves filtering frames of specific data streams and enforcing agreed configurations to guarantee QoS for different data streams. By preventing bandwidth wastage, reducing large data packets, and mitigating harmful or poorly configured endpoints, IEEE~802.1Qci helps ensure reliable and efficient network performance for time-critical applications \cite{Luo2021}. The standard controls how incoming network traffic is managed on each network port. This process involves three steps before data is queued. First, data is divided into different streams, each assigned to a specific gate and meter. Secondly, these gates, which can be either open or closed based on a timed schedule, determine if data can pass. If a gate is open, the data is examined by a flow meter. The flow meter decides if the data can proceed based on specific rules, such as the time since the last data was sent. If the rules are not met, the data is discarded \cite{9348746}. These parts of the standards series enable TSN to provide ultra-high reliability, making it suitable for various applications in smart manufacturing and other domains with critical traffic.

\subsection{Network / Resource Management}
In order to achieve TSN's low latency and highly reliable communication capabilities, proper configuration and management of various network elements, such as bridges, switches, and end devices, as well as resources such as bandwidth, communication paths, and schedules, is crucial. 
One of the base standards for time-sensitive networking, IEEE~802.1Q \cite{8403927}, outlines the methods for managing various categories of traffic on a bridged Ethernet network. The IEEE~802.1Qcc standard \cite{8514112} provides protocol functions for managing and setting up of a TSN network. The IEEE~802.1Qat Stream Reservation Protocol (SRP) guarantees network resources for time-sensitive applications by reserving bandwidth and scheduling between endpoints. This is achieved by allocating resources based on a stream's latency requirements and bandwidth needs, using the Multiple MAC Registration Protocol (MMRP), Multiple VLAN Registration Protocol (MVRP), and Multiple Stream Registration Protocol (MSRP) to manage the reservation process \cite{8467507, 9473542}. Additionally, the IEEE~802.1Qcc and IEEE~802.1Qca standards work together, with IEEE~802.1Qca specifying the abilities for reserving streams of TSN flows, such as path control and bandwidth reservation \cite{8731777}, and IEEE~802.1Qcc bringing enhancements to SRP to manage the streams \cite{9047392}. 

A pertinent standard in resource management is the IEEE~802.1CS Link-local Registration Protocol \cite{9416320}. It outlines methods for duplicating a substantial registration database (approximately 1 MB in size) across a point-to-point connection. IEEE~802.1CS is designed to streamline the creation of application protocols that distribute information across entire or specific network segments \cite{8695835}. This standard addresses the growing need for efficient data management in applications like industrial automation and large-scale audio/video systems, which require extensive configuration and registration information. Unlike IEEE~802.1 MSRP, which is optimized for smaller databases, IEEE~802.1CS excels at handling large-scale data replication, making it a more suitable choice for applications demanding rapid and efficient information dissemination across networks \cite{en14154497}. Furthermore, IEEE~802.1Qcx specifies a system for managing network connectivity faults \cite{9212765}. It describes how network devices like Two-Port MAC Relays, VLAN Bridges, and Provider Bridges should be configured and monitored for problems. This system also includes unified modelling language (UML) diagrams to visualize the network and its components and explains how all these parts work together to manage network faults \cite{10192167}. To safeguard time-sensitive traffic by prioritizing queued frames, the~IEEE~802.1Qci outlines per-stream filtering and policing procedures, which prevent wastage of bandwidth, reduce the occurrence of large data packets, and shield against harmful or poorly configured endpoints \cite{Luo2021}.

\subsection{Challenges for wireless extensions}
Presently, 5G stands as the primary contender for wireless TSN solutions. Engineered to accommodate real-time applications, 5G leverages ultra-reliable and low-latency communication (URLLC) traffic profiles. Through both simulated scenarios and empirical data, it has been established that 5G can deliver worst-case latencies as low as 1 ms, with cycle times ranging from 2 to 3 ms under specific configurations. Consequently, 5G possesses the necessary QoS capabilities for various industrial applications, albeit not for the most demanding ones. 
To support TSN clock synchronization within the 5G radio access network, the 3GPP has recently outlined a procedure to facilitate such synchronization within the 5G system (5GS) \cite{9632397}. 
The second prominent contender for wireless TSN is IEEE~802.11, a widely utilized wireless local area network technology. Within IEEE~802.11, precise clock synchronization is already supported through time/fine time measurement schemes. Furthermore, the most recent IEEE~802.11 amendment, IEEE~802.11ax (Wi-Fi 6), introduces various mechanisms such as the trigger frame, aimed at improving network efficiency and partially enabling time-aware scheduling. Despite these advancements, IEEE~802.11ax performance may still fall short of the desired QoS levels in certain use cases. The forthcoming amendment, IEEE~802.11be (Wi-Fi 7), will incorporate mechanisms such as multilink operation and multi-AP coordination, potentially aiding in reducing worst-case latency \cite{10034532}.

The performance of wireless connections fluctuates depending on the environment (e.g. indoor, factory, open spaces, etc.) and the specific communication parameters (e.g. modulation and coding rate, interference cancellation, etc.). The varying nature of wireless channels and interference from both within and external to a particular system leads to higher higher Packet Error Rates (PER), packet loss rate, latency and jitter \cite{Haxhibeqiri2021}. TSN capabilities, such as bandwidth reservation and scheduling, demonstrably achieve low latency and jitter in Ethernet networks. However, their application to wireless systems necessitates consideration of the inherent characteristics of wireless links, including their time-varying nature of achievable data rates, PER and packet loss rate.

Furthermore, in wireless systems, the need to share a finite medium among numerous devices necessitates the implementation of a well-defined channel access procedure. This is particularly evident in LTE and 5G technology, where delays can originate from both grant acquisition, the process by which devices secure permission to transmit, and random access procedures employed for new device connection. IEEE~802.11's listen-before-talk channel access mechanism introduces a significant source of delay. This approach, also employed by LTE for unlicensed spectrum access, is anticipated to experience further delay increases as the number of devices contending for channel access grows \cite{8469808}. The inherent randomness associated with listen-before-talk protocols presents a challenge for applications demanding stringent real-time performance. Consequently, the development of novel network management methodologies becomes imperative to facilitate ultra-low latency services that require deterministic medium access and minimal jitter \cite{Yang2020}. In this section, we have provided a concise overview of the challenges associated with extending TSN to wireless networks with a more comprehensive analysis following in Section \ref{sec:Open}.


\begin{figure*}[htbp]
\centering
\includegraphics[scale = 0.62] {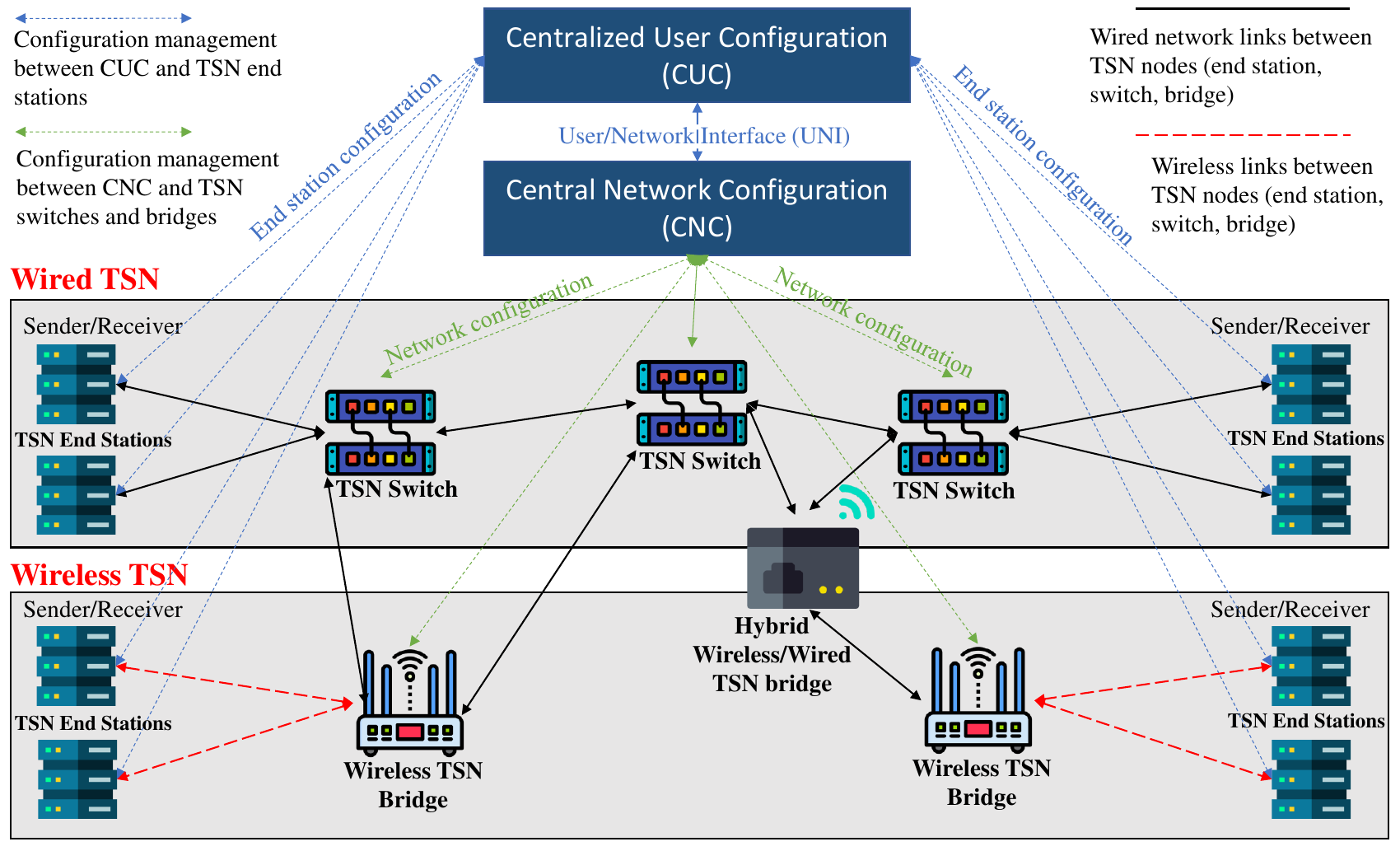}
\caption{TSN architecture overview with wired-wireless-hybrid designs based on IEEE~802.1 Qcc}
\label{overview-arch}
\end{figure*}

\section{Wireless TSN Architecture and R\&D Platforms} \label{sec:Arch}
In this section, an overview of current wireless TSN architectures and their respective components are presented. Following that, a comprehensive discussion of the state-of-the-art, with a particular emphasis on wireless TSN architectures is provided. Lastly, a review of available testbeds and experimental evaluation tools, such as software simulation and hardware implementations, is presented.

\subsection{Architecture overview}

Fig.~\ref{overview-arch} illustrates an overview of a generic TSN architecture with wired-wireless-hybrid design and centralised network management. A TSN network is built with interconnected TSN switches and TSN end stations, also called talkers (for senders) and listeners (for receivers), such as sensors, Programmable Logic Controllers (PLCs), Robots, etc.) linked to these switches. Specific architectures for 5G TSN integration as defined in \cite{3GPP_5GS} and for TSN IEEE~802.11 integration map onto this generic architecture. 

In addition to this, TSN defines models for TSN network management. Three main models are outlined in the IEEE~802.1Qcc amendment: completely centralized, a mix of centralized and distributed management (hybrid), or entirely distributed management \cite{OECHSLE202338}. In the centralized model, the Centralized User Configuration (CUC) and Centralized Network Configuration (CNC) network management components communicate through the User Network Interface (UNI) regarding communication needs specific to users, with CNC managing all the streams and schedules in the network \cite{9483600}. The CUC accepts requests for TSN network services from TSN end station nodes (senders and/or receivers) and transmits them to the CNC, as illustrated by the blue links in the figure. The end stations' setup through an application-dependent configuration protocol falls within the purview of CUC as well. CNC, on the contrary, is fully aware of the network resources and topological data that are accessible. The CNC measures the schedule and establishes the network components in accordance with the specifications that it receives from CUC \cite{9297066}. In the hybrid model, the user's requirements are passed directly to the network, and CNC does the configuration without the existence of a CUC in the network \cite{8368523}. In a fully distributed model, neither CNC nor CUC exist, and the configuration data exchanges between bridges use the YANG model, as specified in IEEE~802.1Qcp \cite{8467507}.

\subsection{State-of-the-art}
Several works have studied or proposed a wired/wireless TSN architecture, which are summarized in Table \ref{Table:Arch}. In \cite{8502524}, a hybrid wired/wireless network consisting of IEEE~802.11 and 5G elements with deterministic characteristics is considered. The considered architecture includes a wired network with TSN and a novel MAC wireless protocol for the wireless part. Additionally, the paper provides information about the developed bridge that enables deterministic data transmission between the two segments of the hybrid network. Several tests have been conducted on real hardware to measure the synchronization accuracy between the hybrid network nodes, the packet delivery determinism of the nodes, and the QoS parameters of the network, such as packet reception latency and wireless node packet error rate.

Authors of \cite{8756136} introduce a novel hybrid wired/wireless architecture design that strives to satisfy the demands for robustness, determinism, and real-time. The suggested design ensures smooth communication between the two media while extending the current wired network into the wireless space. TSN is used to govern the wired network, instead of traditional alternatives such as EtherCAT. The IEEE~802.11 medium access control (MAC) protocol manages the wireless network. An access point (AP) with particular capabilities serving as a deterministic bridge is deployed to improve communication among the two networks. Particularly, the design supports mobility needs for which a soft-handover method is proposed. The proposed technique reduces network overhead and enables continuous communication while avoiding the need for a second radio interface in end devices. The hybrid wired/wireless design enables a broader range of applications and facilitates mobile devices using wireless communications. The proposed hybrid wired/wireless architecture for industrial control applications, while promising, has several notable disadvantages. The reliance on mean RSSI for handover decisions without considering other parameters such as PER or channel statistics may lead to suboptimal performance, and the lack of real-world testing raises concerns about its practical robustness. Scalability issues could arise as the network grows, and the system's heavy dependence on finely-tuned handover parameters may complicate optimal configuration.

In \cite{9968577}, a conversion system architecture that links Operating Technology equipment (industrial automation system devices) and engages with edge devices is presented. Within this architecture, a hybrid wired/wireless protocol conversion module is implemented to translate various industrial protocols and integrate them with the Ethernet network. The authors introduce a Flexible Protocol Conversion Mechanism that distinguishes between the data plane and the configuration plane. Moreover, a TSN-compatible frame structure with an adaptive priority mapping mechanism that modifies the priority in the VLAN tag according to the delay in the computation is developed. Based on the experimental results, the conversion latency of each tested pair of protocols is consistently less than 250 milliseconds. For instance, in PROFINET and EtherCAT, the Min and Max values range from 60 to 115 milliseconds.

In \cite{9447967}, a hybrid wired/wireless high-precision time synchronization network based on a deterministic TSN called Time-triggered Ethernet (TTE) with the ability to provide stable latency and minimal jitter is presented.  TTE has the ability to correct transmission jitter produced throughout traffic propagation and switch functioning. TTE is regarded as the primary backbone network, while 5G-URLLC-based wireless access is employed as a sub-network and assisted by the PTP. The simulation outcomes illustrate how the integration of PTP-aided URLLC effectively preserves network reliability, latency, and jitter while operating seamlessly with the TTE network.

In \cite{10077954}, a hierarchical computing network cooperation framework for IIoT that determines the best forwarding route and time by effectively gathering and coordinating time-critical traffic necessities, computing resources, and current network state in the network hierarchy is presented. In order to achieve infrastructure connectivity and flexible resource adaption in the highly dynamic environment of IIoT, a modified TSN scheduling shaper method is utilized. The proposed approach involves partitioning the network topology into separate domains known as Computing Network Collaboration Domains (CNCDs). This division allows for more effective management of the network while ensuring the deterministic delay in network link transmission by controlling the forwarding time of each task data traffic frame through a Gate Control List (GCL) within a priority queue. Additionally, the proposed method intelligently matches the available distributed computing resources to optimize computing and network integration, resulting in enhanced analysis and processing speed for time-sensitive tasks in industrial settings. The effectiveness of the prototype system is assessed through Omnet++ based simulation, and the outcomes suggest that the presented architecture shows promise in meeting the demanding criteria of low latency, minimal jitter, and exceptional reliability for extensive industrial applications, though it may not fully achieve these goals under all conditions. Reliance on the Omnet++ simulation may limit real-world applicability, and using a genetic algorithm in C++ could add computational overhead, reducing efficiency gains. The chosen evaluation metrics may not fully capture system stability and scalability in dynamic environments. The test scenarios lack extreme cases, potentially underestimating weaknesses. The comparison with only traditional models may overlook newer hybrid and adaptive systems, missing a more rigorous evaluation. 

In \cite{telecom5010004}, a Time-Sensitive-Enabled Small Cell (TESC) architecture is introduced to enhance synchronization precision for industrial IoT (IIoT) devices. The main contribution is the introduction of a Small Cell TSN Translator (SC-TT) to offload synchronization tasks from UE, improving synchronization accuracy to a range between 0.01 \(\mu\)s and 1 \(\mu\)s. Through simulations, the authors demonstrate that using the TESC can achieve high-precision synchronization (down to 10 \(\mu\)s) while efficiently managing Resource Block (RB) consumption, making it suitable for applications requiring precise timing, such as autonomous vehicles and industrial automation. The paper primarily focuses on simulations to demonstrate the effectiveness of the TESC architecture, but it lacks discussion on the challenges and feasibility of real-world implementation. Key practical factors, such as the potential for interference in dynamic industrial environments and the reliability of synchronization in varying conditions, such as high mobility or network congestion, are not fully explored.

\begin{table}[htbp]
\caption{Summary of existing work on wireless TSN architecture}  \label{Table:Arch}    
\scriptsize
\centering          
\begin{tabular}{|c|c|c|c|c|c|}  
\hline 
\textbf{Work} & \textbf{5G} & \textbf{802.11} & \textbf{Hybrid design} & \textbf{Simulation/Experiment} \\[0.5ex]  
\hline \hline
\cite{8368523} & \checkmark & \checkmark & \checkmark  & Software evaluation \\ \hline
\cite{8502524} & \xmark & \checkmark & \checkmark  & Hardware implementation \\ \hline
\cite{8756136} & \xmark & \checkmark & \checkmark  & Software evaluation \\ \hline
\cite{9968577} & \xmark & \checkmark & \checkmark & Hardware implementation \\ \hline
\cite{9447967} & \checkmark & \xmark & \checkmark & Software evaluation \\ \hline
\cite{10077954} & \xmark & \checkmark & \checkmark  & Software evaluation \\ \hline
\cite{9632397} & \checkmark & \checkmark & \checkmark & Hardware implementation \\ \hline
\cite{9212141} & \checkmark & \xmark & \xmark & Hardware implementation \\ \hline
\cite{9272460} & \checkmark & \xmark & \xmark  & Hardware implementation \\ \hline
\cite{telecom5010004} & \checkmark & \xmark & \xmark  & Software evaluation \\ \hline

\end{tabular}
\end{table}

\subsection{Testbeds and experimental implementations}
Wireless TSN experimental implementations and testbeds are currently scarce despite the growing interest in the topic. In several works, including \cite{9625170, 9779191}, the OMNeT++ simulation platform \cite{Varga2010} with a TSN framework, known as NeSTiNg (Network Simulator for Time-Sensitive Networking) \cite{8854500}, has been used. OMNeT++ is an open-source, fully accessible and extensible freely available network simulation platform widely used in wired and wireless network research. OMNeT++ promotes contributions by researchers to its frameworks to extend its capabilities towards a wide range of networking technologies and provide them to the research community. The core purpose of NeSTiNg framework was to establish the foundational TSN elements within OMNET++. Following this, the INET framework models were employed to expand the TSN capabilities of NeSTiNg from the wired setup into the wireless domain \cite{9698580}. Recent investigations such as \cite{10333533, electronics13040768} have leveraged the iNet and Simu5G frameworks to comprehensively evaluate the performance of 5G-TSN architectures. These studies offer valuable insights into the feasibility and potential benefits of integrating TSN functionalities into 5G networks.

On the hardware testbed side, various research projects utilized distinct networking hardware components within their respective testbeds for a physical implementation of a wireless TSN prototype. In \cite{10060159}, an updated summary of research in the field of 5G-TSN, covering simulation tools, the open-source 5G System, and the hardware options available for integrating these components, are presented. In \cite{9272460}, the EchoRing evaluation kit is used to assess the guaranteed latency, and Wireshark is used to record mirrored traffic on a measuring system. In \cite{9453686}, Openwifi access points with beacon follow-up packets enabled and COTS hardware clients were used for the evaluations. In \cite{9779186}, a feasible wireless TSN association method using beacons, developed and tested under real-world conditions on an FPGA-based Software-Defined Radio (SDR) platform utilizing the IEEE~802.11 network. The SDR ADRV9361-Z7035, which integrates the Xilinx Z7035 Zynq®-7000 with the Analog Devices AD9361 integrated RF Agile Transceiver, was equipped with the Openwifi IEEE~802.11 baseband chip/FPGA design in order to evaluate the suggested solution. Additionally, the carrier card ADRVICRR-BOBIFMC was used to add an Ethernet connection. In \cite{10634074}, the authors focused on implementing and testing two key features of the upcoming IEEE~802.11be standard: R-TWT and C-SR. They integrated these features into the Openwifi platform. Their main goals were to assess how R-TWT can improve latency for time-sensitive applications and how C-SR can enhance overall network throughput, particularly in dense network environments with multiple APs. The results of their assessment suggest that R-TWT has the potential to reduce latency for time-sensitive applications, possibly meeting required latency thresholds. Additionally, C-SR could improve overall network throughput by up to 50\% for UDP traffic and up to 960\% for TCP traffic in congested environments. However, the implementation of R-TWT and C-SR may face challenges in real-world deployment, particularly in maintaining synchronization and managing scheduling overhead. Additionally, the region-based C-SR approach might limit granularity in selecting concurrent transmissions, potentially reducing the overall efficiency in highly dense networks. In \cite{10209171}, a testbed for TSN over 5G is introduced. The testbed consisted of two main TSN nodes from Relyum, a Telit fn980m modem as the 5G UE, two TSN translators that consist of one device-side TSN translator and one network-side TSN translator, one TSN bridge utilized as CNC and CUC, Open5GS to deploy 5G core and ADVA devices for time synchronization. 

In \cite{10637136}, the authors investigated the performance of OFDMA in Wi-Fi 6 for latency-sensitive industrial control applications, specifically the leader-follower dual-lift robotic use case. They aimed to determine whether OFDMA could provide the low latency and reliability needed for real-time control. Using a testbed, they measured the round trip time (RTT) and Cartesian error between leader and follower robots under different network conditions, with and without OFDMA enabled. Contrary to expectations, they found that enabling OFDMA increased latency and jitter, leading to degraded performance compared to traditional CSMA and wireless TSN. OFDMA performed better when moderate background traffic was introduced, keeping the OFDMA trigger active. However, even with doubled channel bandwidth, OFDMA failed to meet the strict latency requirements of the application, while wireless TSN showed more deterministic performance. The study concluded that improvements to OFDMA scheduling are needed for real-time industrial applications and recommended future refinements in AP implementations and the IEEE~802.11 standard. In \cite{10443947} the researchers aimed to integrate mobility and TSN by developing a novel method for seamless handover in Wi-Fi-based wireless TSN. They tested various approaches to optimize the selection of handover moments and reduce delays during handovers. Their results demonstrated that combining Received Signal Strength Indicator (RSSI) and location-based online learning could eliminate the need for traditional scanning methods, significantly improving the handover process. Their method was found to be more effective than offline learning and midpoint selection techniques, achieving a handover delay of less than 20 milliseconds. This achievement enables a more seamless transition between APs, making it particularly beneficial for time-sensitive applications, such as extended reality (XR). The research highlights the potential for this approach to revolutionize technologies reliant on low-latency wireless communication. However, the approach may struggle with real-time adaptability in highly dynamic environments where rapid changes in network conditions or interference occur. 

\subsection{Summary and lesson learned} \label{sec:SummarySec3}
While the 3rd Generation Partnership Project (3GPP) has standardised the architecture of how 5G can be used as a wireless bridge in TSN (see Figure \ref{fig3}), the integration of IEEE~802.11/Wi-Fi is much less clear. A further open issue is how the network elements of a wireless TSN network can be configured. The integration of the 5G control plane with the CNC/CUC management model of TSN is still open as is how the management of Wi-Fi access points fits into the three management architectures of TSN. Research has therefore focused on developing hybrid wired/wireless architectures. These designs often leverage TSN for wired networks and IEEE~802.11 or 5G for wireless networks, with specialized mechanisms for seamless integration and communication between the two media.

Experimental evaluations and testbed implementations play a key role in validating and assessing the proposed architectures' performance. While simulation tools such as OMNeT++ with TSN frameworks provide a versatile platform for software evaluation and studying scalability, hardware implementations using networking components offer more realistic insights into real-world performance and deployment scenarios. In particular, the simulation tools suffer from too simplistic wireless communication models, which affects the accuracy of the performance evaluation. Overall, the lessons learned emphasize the importance of devising flexible and adaptive architectures, more accurate simulation models and experimental implementations of recent standards and their evaluation to devise real-world solutions for wireless TSN.

\section{Wireless Technologies for TSN} \label{sec:Tech}

With the emergence of wireless communication technologies such as 5G and IEEE~802.11ax, TSN has been expanded to include wireless networks, offering flexible installation options compared to wired networks, easier deployment, mobility, and cost-effective maintenance \cite{9134382}. The IEEE~802.11 working group and the 3rd Generation Partnership Project (3GPP) are the primary organisations that work on extending wireless standards to include TSN capabilities. the IEEE~802.11 WG has added relevant capabilities to the IEEE~802.11ax and IEEE~802.11be standards, whereas 3GPP have introduce TSN capabilities into 5G starting with Release-16 of the standard \cite{9855453}. 
With the rise of novel applications in consumer, industrial, transportation, and healthcare domains, wireless technology needs to meet increasingly demanding performance requirements. Each new generation of wireless technology strives to address the limitations of its predecessors and offer advancements in performance, effectiveness and efficiency. This ongoing development has been instrumental in delivering reliable wireless connectivity to consumers and businesses.

\subsection{Taxonomy: Wireless Time-Sensitive Networking}
For wireless Time-Sensitive Networking (TSN), both the wireless and TSN aspects encompass a variety of technologies and standards. To categorize the research areas based on these different technologies and standards, we have developed a taxonomy, illustrated in Table~\ref{Table:WTSN-wireless-5g}. The first level of categorization is based on the wireless technologies, which include Wi-Fi, 5G, and 6G. Each technology is then further divided based on its generation or release version. For example, the 5G section first illustrates the presented works on Release 16, followed by Release 17. The third level of classification further refines the categorization based on TSN standards. Here, the criteria for differentiation follow the structure outlined in Fig. \ref{fig_tsn}. These criteria include factors like time-synchronization, latency, reliability, and resource/network management. 


\subsection{Wi-Fi based TSN}
In Wi-Fi, the predominant deployments currently rely on IEEE~802.11n (Wi-Fi 4) and IEEE~802.11ac (Wi-Fi 5) standards. The combination of Multiple Input Multiple Output (MIMO) and Spatial Multiplexing enhances data throughput by exploiting multipath propagation. Furthermore, Frame Aggregation and Short Guard Interval (GI) decrease the overhead while increasing the efficiency. To improve reliability and performance in noisy environments, error-correcting codes, such as LDPC (Low-Density Parity-Check) and STBC (Space-Time Block Coding) are offered. IEEE~802.11ac achieves significant performance improvements through the use of several key features: wider channels (80 MHz and 160 MHz), enhanced frame aggregation, Quadrature Amplitude Modulation (256-QAM), and Multi-User MIMO, which enables simultaneous data transmission to multiple devices. 

The next significant update is IEEE~802.11ax (Wi-Fi 6), which allows for simultaneous multi-user transmissions, providing the potential to reduce latency and enhance reliability. Moreover, the target wake time (TWT) feature is utilized by STAs to establish a wake time schedule that prompts them to awaken periodically for the purpose of transmitting or receiving data. Furthermore, IEEE~802.11ax introduces more efficient radio resource management using Orthogonal Frequency Division Multiple Access (OFDMA), increased data density utilizing 1024-QAM, and improved signal targeting with the help of Spatial Reuse to deliver higher data rates and handle more devices simultaneously \cite{en14154497}. An access point can transfer data to multiple stations at once using OFDMA, and the other way around. Given that several data transmissions use a similar PHY header and channel access timeout intervals, OFDMA is particularly effective for many brief broadcasts across wide channels. Additionally, OFDMA improves spectral power density during uplink transmission and makes improved fading management possible \cite{8422767}.

The imminent IEEE~802.11be amendment (known as Wi-Fi 7), which follows IEEE~802.11ax features and technologies such as enhanced OFDMA, Enhanced Distributed Channel Access (EDCA), distributed MIMO, multi-AP coordination, and low-latency improvements, which have the potential to lower the worst-case latency \cite{9789864, 9090146}. In addition, it introduces Coordinated Spatial Reuse (C-SR) and Restricted Target Wake Time (R-TWT) as major innovations to enhance spectrum efficiency, throughput, and support for time-sensitive applications, marking a significant step forward in Wi-Fi technology evolution \cite{10634074}. In the following, we provide a brief overview of the research conducted on Wi-Fi based TSN technology.

\begin{table*}[htbp]
\caption{Summary of key existing work on wireless technologies for TSN}  \label{Table:WTSN-wireless-5g}    
\scriptsize
\centering          
\begin{tabular}{|c|c|c|c|c|c|c}  
\hline 
\textbf{Wireless Technology} & \textbf{Standard/Release}      & \textbf{TSN Standard}                     & \textbf{Focus Area}                                 & \textbf{Performance Metric}                                                                           & \textbf{Simulation/Experiment}                                    & \textbf{Work}             \\ \hline
\multirow{11}{*}{Wi-Fi}      & \multirow{6}{*}{IEEE~802.11}   & \multirow{3}{*}{IEEE~802.1AS}             & \multirow{3}{*}{Time Sync}                          & Overhead                                                                                              & Openwifi AP, PTP over Wi-Fi                                       & \cite{9453686}            \\ \cline{5-7} 
                             &                                &                                           &                                                     & synchronization error                                                                                 & \makecell{FPGA-Based \\ Hardware implementation}                  & \cite{9169833}            \\ \cline{5-7} 
                             &                                &                                           &                                                     & Clock offset                                                                                          & Simulation                                                        & \cite{9145977}            \\ \cline{3-7} 
                             &                                & IEEE~802.1 Qbv                            & \makecell{Time Sync, \\ traffic scheduling}         & \makecell{Timestamp delay \\ arrival offsets}                                                         & \makecell{Openwifi \\ IEEES02.11/Wi-Fi baseband \\ chip/FPGA}     & \cite{9779186}            \\ \cline{3-7} 
                             &                                & \multirow{2}{*}{--}                       & Latency, reliability                                & PER, latency                                                                                          & Hardware implementation                                           & \cite{2022-3901}          \\ \cline{4-7} 
                             &                                &                                           & Scheduling                                          & \makecell{Delay, PER \\ Throughput}                                                                   & OMNeT++                                                           & \cite{9625170}            \\ \cline{2-7} 
                             & IEEE~802.11v                   & \makecell{IEEE~802.1AS \\ IEEE~802.1CB}   & Latency, reliability                                & Clock offset, reliability, QoS                                                                        & --                                                                & \cite{info12010012}       \\ \cline{2-7} 
                             & IEEE~802.11ad                  & \makecell{IEEE~802.1CB}                & Delay, reliability                                  & \makecell{No. of APs,\\ Load distribution}                                                            & Matlab                                                            & \cite{GARROPPO2023109771} \\ \cline{2-7} 
                             & IEEE~802.11ax                  & --                                        & \makecell{control cycles, latency, \\ reliability } & PER, communication jitter, latency                                                                    & Hardware implementation                                           & \cite{9134382}            \\ \cline{2-7} 
                             & \multirow{2}{*}{IEEE~802.11be} & --                                        & Reliability, latency                                & PER, latency                                                                                          & Matlab                                                            & \cite{9557495}            \\ \cline{3-7} 
                             &                                & --                                        & Transmission scheduling                             & Throughput, latency                                                                                   & Matlab                                                            & \cite{10001702}           \\ \hline
Wi-Fi/5G                     & --                             & IEEE~802.1AS                              & Time Sync                                           & --                                                                                                    & Hardware implementation                                           & \cite{9443078}            \\ \hline
\multirow{12}{*}{5G}         & \multirow{5}{*}{Release 16}    & --                                        & Time Sync                                           & Generic                                                                                               & --                                                                & \cite{9204594}            \\ \cline{3-7} 
                             &                                & --                                        & QoS Mapping, scheduling                             & \makecell{Latency,  resource utilization \\ Throughput }                                              & Hardware implementation                                           & \cite{10110348}           \\ \cline{3-7} 
                             &                                & --                                        & Resource allocation                                 & No. Supported TSC UEs , CDF                                                                           & Simulation                                                        & \cite{9138407}            \\ \cline{3-7}
                             &                                & --                                        & Resource allocation                                 & \makecell{Convergence performance \\ Resource utilization}                                                                           & Simulation                                                        & \cite{10623103}            \\ \cline{3-7}                             
                            &                                & --                                        & Resource allocation                                 & \makecell{Reliability, runtime\\ resource utilization }                                                                          & Matlab                                                        & \cite{10637302}            \\ \cline{3-7}
                             &                                & 802.1Q/TCI                                & Generic                                             & End-to-end delay                                                                                      & OMNeT ++ with NeSTiNg                                             & \cite{9779191}            \\ \cline{3-7} 
                             &                                & \makecell{IEEE~802.1Qbu \\ IEEE~802.1Qbv} & --                                                  & Latency, jitter                                                                                       & \makecell{TSN Ethernet switches \\ Ixia traffic generator, iperf} & \cite{9149161}            \\ \cline{2-7} 
                             & Release 17                     & IEEE~802.1AS                              & Time Sync                                           & Synchronization errors                                                                                & NS-3                                                              & \cite{9557468}            \\ \cline{2-7} 
                             & \multirow{6}{*}{--}            & Generic                                   & Time Sync                                           & Throughput, jitter                                                                                    & Hardware implementation                                           & \cite{9785739}            \\ \cline{3-7} 
                             &                                & --                                        & Latency                                             & CDF                                                                                                   & Hardware implementation                                           & \cite{9779167}            \\ \cline{3-7} 
                             &                                & --                                        & Latency, reliability                                & \makecell{Cumulative Distribution \\ Function (CDF), \\ latency}                                      & \makecell{Prototype system with \\ 5G NR system emulator}         & \cite{8643425}            \\ \cline{3-7} 
                             &                                & --                                        & Latency, reliability                                & \makecell{Response time, packet loss rate, \\ throughput}                                             & Telematics simulation                                             & \cite{9828687}            \\ \cline{3-7} 
                             &                                & --                                        & Transmission scheduling                             & Latency, standard deviation                                                                           & Hardware implementation                                           & \cite{10062219}           \\ \cline{3-7} 
                             &                                & --                                        & End-to-End Scheduling                               & \makecell{Scheduling success ratios \\ resource utilization, }                                         & Software simulation                                               & \cite{10136614}           \\ \hline
\multirow{2}{*}{6G}          & \multirow{2}{*}{--}            & IEEE~802.1Qcc                             & Latency, reliability                                & --                                                                                                    & --                                                                & \cite{10275568}           \\ \cline{3-7} 
                             &                                & -- & Latency, reliability & \makecell{Convergence performance,\\ Average scheduling success \&\\ failed ratio, testing accuracy } & Simulation& \cite{10210522}           \\ \hline
\end{tabular}
\end{table*}

\subsubsection{Time-synchronization}

In the study presented in \cite{9453686}, the authors investigated Wi-Fi based TSN and proposed a low-overhead beacon-based time synchronization mechanism. This mechanism achieves an average synchronization accuracy of as low as 10 microseconds for a low beacon interval. It includes a follow-up beacon packet to account for delays in sending beacons and was tested in various intra and inter-access point communication settings. The results indicate that the synchronization performance depended on the synchronization accuracy of the access points in the intra-AP setting. Furthermore, the impact of beacon interval and wireless link load on synchronization was validated. The proposed mechanism was benchmarked against the performance of PTP over wireless links. To reduce overhead and enhance synchronization accuracy, further improvements are possible, such as the ability of the physical layer to directly timestamp the beacon packet, eliminating the need for a follow-up packet. In \cite{9169833}, an innovative mechanism for precise synchronization of wireless networks is presented, aiming to extend TSN to IEEE~802.11 networks. This mechanism utilizes the time synchronization function (TSF) with a master clock as the source. By modifying the controlling software, it maximizes the potential of low-cost commercial wireless chips. Consequently, field-programmable gate arrays (FPGAs) required for timestamping data from the wireless chip were simple and uncomplicated. Unlike relying solely on software, this approach avoids utilizing the resources of the embedded processor hosting the wireless chip, thus mitigating potential issues like additional and unpredictable delays associated with CPU usage. The mechanism achieves synchronization errors with a standard deviation of less than 500 ns, even under varying CPU and network loads, comparable to contemporary wired real-time field buses.

In \cite{9443078}, a comprehensive examination of various wireless clock-synchronization methods and their achievable performance is provided, while also exploring their potential for facilitating wireless TSN. The study concludes that the current clock-synchronization techniques are adequate for enabling wireless TSN. However, it is emphasized that substantial implementation endeavours will be necessary to integrate precise clock synchronization into wireless systems.
In \cite{9145977}, a WLAN-based mechanism is presented to satisfy requirements such as clock synchronization and spatial dynamics of clients in wireless networks. In this study, a mechanism aimed at utilizing PTP in wireless networks, leveraging the characteristics of the medium to improve performance is proposed. Additionally, a modification to the protocol's precision-enhancing segment to minimize communication overhead is introduced. The solution ensures the retention of the minimum necessary number of transmitted messages while achieving improved synchronization precision.

\subsubsection{Latency and reliability}

Authors in \cite{9134382} introduced the wireless SHARP (w-SHARP) hardware implementation, which offers time synchronization, time-aware scheduling with constrained latency, and excellent reliability. This implementation utilizes a software-defined radio architecture based on field-programmable gate arrays. The w-SHARP technology demonstrates minimal control cycles, low latency, and high reliability using a hardware testbed, making it a potential wireless solution for real-time industrial applications.

A TDMA-based protocol called RT-WiFiQA for IEEE~802.11 networks, aiming to enhance reliability and latency performance, was proposed in \cite{2022-3901}. This protocol utilizes two distinctive approaches: real-time quality of service (RT-QoS) and fine-grained aggregation (FGA). The RT-QoS protocol ensures the quality of service requirements for various traffic types, particularly in the presence of diverse traffic classes, while supporting the FGA mechanism. The FGA mechanism aggregates packets from multiple stations, minimizing transmission overhead to improve network efficiency and reliability. The study introduces a crucial limit that determines when the FGA method outperforms the non-aggregation approach based on the trade-off between the size of the FGA packet and its reliability. The protocol was implemented on commercially available hardware running the Linux OS and was evaluated through experiments comparing its performance with RT-WiFi and traditional Wi-Fi. The results demonstrate that RT-WiFiQA exhibits greater reliability than RT-WiFi and performs real-time functions more effectively than Wi-Fi at both the MAC and application layers. However, there were limitations, including the reliance on a fixed rate control mechanism for each packet transmission, which restricts the protocol's reliability. Additionally, the existing RT-WiFiQA protocol could not be extended to more contemporary OFDMA IEEE~802.11 systems such as IEEE~802.11ax due to hardware constraints. It was designed for usage with commercially available chips employing an OFDM physical layer, specifically IEEE~802.11 b/g/n.

In another relevant study \cite{GARROPPO2023109771}, a novel design framework for IEEE~802.11ad networks for TSN-based industrial applications is presented. The paper focuses on two key aspects concerning the implementation of IEEE~802.11ad technology in an industrial setting. The first aspect is to ensure reliable data delivery and minimize latency by optimizing the parameter settings of the IEEE~802.11ad MAC through the utilization of the synchronous service periods (SPs) mechanism. The second aspect involves proposing a binary linear programming model to improve system reliability using the Frame Replication and Elimination for Reliability (FRER) technique. The number of deployed APs is minimized in the model, considering the limitations imposed by the available synchronous service periods (SPs). Additionally, two heuristics are introduced based on the mathematical model described, aiming to address the design problem effectively. Furthermore, a comparison is made to evaluate the computational performance of the proposed approaches.

The authors in \cite{9557495} evaluate the Multi-Link Operation (MLO) approach for the IEEE~802.11be standard, specifically focusing on its applicability in industrial settings with latency-bound Wireless TSN requirements. Simulations were conducted using data obtained from a radio measuring experiment conducted by the National Institute of Standards and Technology (NIST) at manufacturing facilities. Two wireless networks with distinct characteristics were simulated, and parameters such as Packet Error Rate (PER) and different fading channel models (Rayleigh and Rician) were compared to assess these channels. The simulations were implemented using custom MATLAB code with support from the MathWorks WLAN Toolbox. The results of the simulations demonstrated that significant improvements in latency and reliability can be achieved in practical industrial scenarios with interference limitations by employing MLO in conjunction with the IEEE~802.11ax standard. Specifically, an improvement of over 3 dB was observed for a latency target of 1 ms and a success rate of 99.9\%. The primary factor contributing to the latency improvement was a reduction in the number of re-transmissions required for successful frame transmission. However, to evaluate the Multi-Link capabilities in real-world industrial applications, it is necessary to implement MLO in a hardware testbed, which was not undertaken in the study.

In \cite{info12010012}, presents a brief overview of the key functionalities that define a wireless TSN system and examines the approaches that have been explored in the literature and in standardization efforts. This is followed by the description of a framework for software-based implementation, a toolbox that enables one to integrate the necessary capabilities into a practical solution in wireless TSN. Additionally, some design aspects that will enhance transmission reliability, one of the major obstacles in wireless communication, to attain a performance equal to wired systems are discussed, are discussed.

\subsubsection{Traffic scheduling and resource allocation}

In \cite{9625170}, a novel channel access method for MAC protocol that uses a per-flow TAS scheduler to schedule sensing slots using a collision-free CSMA protocol called Transmission Gating Time Hyperchannel (TGT-HC) is presented. To improve its overall performance, a combination of TAS and First-In-First-Out (FIFO) scheduling is employed to ascertain the sensing slot assignments. Based on the obtained results, it is determined that this upgraded protocol exhibits a delay performance that closely resembles that of an ideal First-Come-First-Serve (FCFS) single server when dealing with cyclic traffic. While currently serving as an independent solution, there is potential for the TAS-driven contention-free CSMA scheme to be integrated into forthcoming TSN enhancements for Wi-Fi 7 and future generations. 

In \cite{10001702}, the concept of Scheduled Transmission Opportunities (S-TXOP) is illustrated, in which the network controller utilizes the predictable traffic patterns related to time-sensitive applications to reduce control overhead and offer scheduled channel access, achieving significantly low latency cycles. According to the simulation results, the S-TXOP outperforms current scheduling methods and demonstrates higher robustness and efficiency in addressing these challenges used in IEEE~802.11.

In \cite{9779186}, a wireless TSN impactless association mechanism is presented, employing beacons to synchronize time and schedule traffic for future wireless TSN clients during the association stage. In addition to providing the association schedule, this standard-compliant solution improves beacon-based synchronization by filtering delayed signals. By removing the possibility of traffic collision among potential wireless TSN users and those already linked, this method improves the overall determinism of the network. The unique association architecture was developed and evaluated using the Openwifi 802.11 software in practical settings over a wireless Software Defined Radio platform. The extensive evaluation demonstrates that this mechanism achieved high-accuracy synchronization, even with complex scheduling timeslots, such as short timeslots of 128 microseconds.

\subsection{5G-based TSN}
\begin{figure*}[t]
\centering
\includegraphics[width=5.25 in, height=2.5 in] {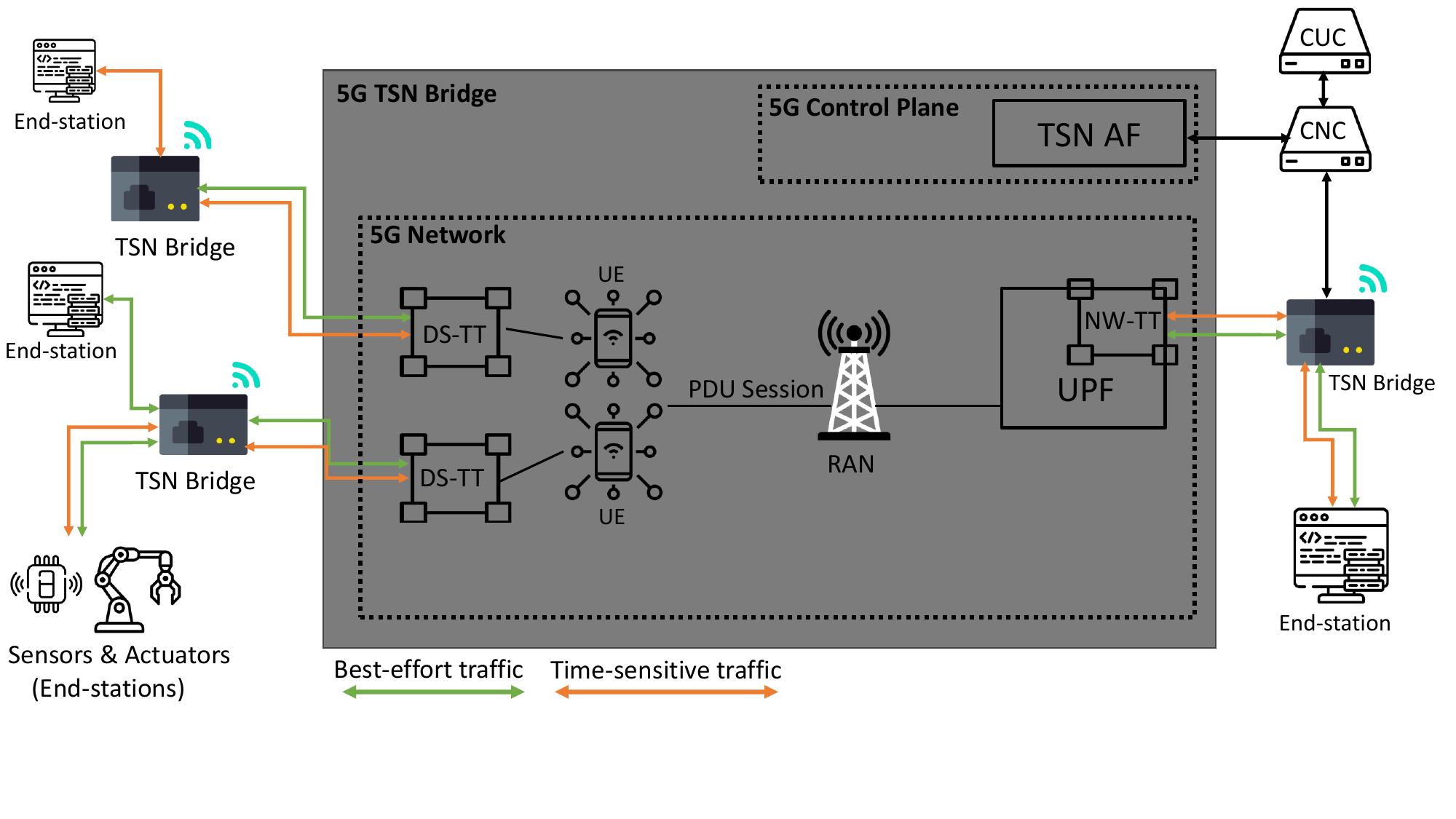}
\caption{Overview of the 5G logical TSN bridge}
\label{fig3}
\end{figure*}

Within the 3GPP standardisation framework, the idea of integrating the 5G network as a logical TSN bridge stands out as the most favourable strategy for enabling TSN functionality over 5G \cite{9921731}. The fundamental concept involves establishing a black box model wherein the 5G system acts as a TSN bridge for the external TSN network. However, internally, it employs its own infrastructure to transmit TSN frames \cite{9065178}. Fig. \ref{fig3} illustrates the overview of 5G-TSN in which the 5G system serves as a TSN bridge, transporting TSN streams consisting of both best-effort and time-sensitive traffic within its internal 5G QoS framework. Two architectural strategies are offered by 3GPP for utilizing 5G as a logical TSN bridge \cite{3gpp2019, 3gpp2023}. 

In the first architecture, the TSN translator (TT) resides inside the User Plane Function (UPF). The TSN Translator and the UPF are considered as a single component in this design, with the UPF serving as the hub for parameter translation between TSN and 5G \cite{9779191}. In the second architecture, the TSN translators are positioned on the device-side (DS-TT) and the network-side (NW-TT) of the logical TSN bridge. These translators perform traffic scheduling by utilizing hold-and-forward mechanisms from IEEE~802.1Q standards to reduce jitter in data flows. The role of the NW-TT includes translating 5G QoS to TSN QoS, managing traffic at the core network edge, and ensuring that time-sensitive data maintains its integrity and priority. Meanwhile, the role of the DS-TT involves translating TSN parameters to 5G QoS on the device side, interfacing with the UE, and establishing message priority for proper handling within the 5G network \cite{10541017}. The TSN Adapter Function (TSN AF) is responsible for mediating the QoS mapping between TSN traffic classes and their corresponding 5G QoS profiles. This mapping process leverages configuration information received from CNC. In essence, the CNC dictates the desired TSN QoS parameters, such as priority and delay, for various traffic classes. The TSN AF then translates these directives into concrete configurations within the 5G network, ensuring that data is prioritized and delivered according to the pre-defined latency requirements \cite{9212141}. With these key components working in concert, the 5G network establishes a robust foundation for deterministic communication to act as a transparent wireless TSN bridge.

3GPP has undergone significant milestones in developing and deploying enhanced versions of the 5G standard in recent years. The main objective of 3GPP Release 15 was the development of enhanced Mobile Broadband (eMBB), while 3GPP Release 16 aimed to provide additional functionalities for Industrial IoT (IIoT). 3GPP Release 15 also introduced URLLC to support mission-critical applications requiring extremely low latency and high reliability. This technology, combined with 5G New Radio, Enhanced Mobile Broadband (eMBB), and beamforming, enables both higher data rates and lower latency. It also includes other technologies such as enhanced Location Services, support for Non-Public Networks (NPNs), and 5G LAN services. It establishes the structure and specifications for TSN translation between 5G systems (5GS) and TSN-capable networks. Furthermore, 3GPP Release 16 introduces Ethernet Protocol Data Unit (PDU) sessions that allow bridging between the Ethernet MAC layer and 5G. 3GPP Release 16 brought several enhancements, particularly supporting TSN in the context of 5G networks with URLLC, network slicing, and Massive Machine Type Communications \cite{9921680}. As part of the next 5G release, Release 17, the objective of 5GS is to enable uplink synchronization for compatibility with TSN. In Release 16, only the TSN GM connected to the network side was compatible. However, in Release 17, there is a plan to create a standard technology in which 5GS will offer time synchronization when multiple TSN GMs are connected to the user equipment (UE) end. Moreover, Release 17 has suggested several approaches to facilitate UE-to-UE connectivity, because uplink synchronization will automatically occur \cite{Fetrij.2020-0200}

\subsubsection{Time-synchronization}

The IEEE~802.1AS standard does not explicitly address TSN time synchronization over 5G. Instead, the 3GPP specification TS 23.501 outlines the functional requirements and procedures. To enable TSN-compliant time synchronization, a 5G network must emulate a regular PTP relay as defined by IEEE~802.1AS \cite{9562179}. This means that the 5G network should appear as a standard switch to other network components. TSN devices at the edge of the 5G network participate in time synchronization processes similarly to how they would in a traditional Ethernet-based network. While the 5G network must remain transparent to the TSN network, it requires additional mechanisms compared to an Ethernet-based PTP relay. Notably, TTs are introduced at the edge of the User Plane and the UE, necessitating the division of the time synchronization implementation into two components: NW-TT and DS-TT. Furthermore, additional configuration steps are required, such as configuring the user plane to prioritize time synchronization over TSN stream transportation \cite{10541017}. The 5G TSN AF enables time synchronization support on both NW-TT and DS-TT when requested. To fully comply with IEEE~802.1AS, the 5G network must adhere to the maximum residence time requirement. The 5G network can theoretically verify this by examining the UE-provided DS-TT residence time and the assigned PDU Flow Identifier (QFI). If this information is unavailable, a default value can be used. However, this poses a challenge as the Ethernet ports of the virtual switch may be physically separated, with one behind the UPF and the other connected to the UE. Timing information, in the form of gPTP messages, can originate from various locations, including the NW-TT, DS-TT, or even the 5G network itself. In scenarios with UE-to-UE traffic, PTP messages must still traverse the UPF, and the Best Master Clock Algorithm continues to operate in the NW-TT \cite{9674640}.

Authors in \cite{9204594} demonstrated how 3GPP TS 23.501 Release 16 provides the necessary foundation for time synchronization in industrial automation using 5G networks. The convergence of 5G networks with wired networks that will both be based on TSN in the future is described. Furthermore, a number of factors that must be taken into account to achieve precise time synchronization between 5G and TSN in integrated deployments are discussed. In \cite{9785739}, an innovative mechanism for zero touch self-configuration of private 5G networks to carry TSN traffic is introduced. The capacity to explore various setups for given traffic and anticipated Key Performance Indicators (KPIs) using automated Model-Based Testing is one of the advantages of employing automata learning to track and analyze the actions of networks as state machines. The paper concentrates on the foundational step of the entire procedure, which is building a model of the 5G network using automata learning. The algorithm has been verified in a testbed using actual traces to enable TSN over 5G.

In \cite{9557468}, the effects of synchronization and synchronization faults on the possible end-to-end time synchronization precision for 5G UE-to-UE communication in the 5GS using integrated 5G and TSN networks have been studied. Based on 3GPP Release 17, the study focuses on the effects of the 5G System emulating a TSN transparent clock. Initially, the evaluation is focused on synchronization errors pertaining to the 5G ingress and egress devices. Then, the analysis is expanded to encompass synchronization errors occurring between these 5G ingress and egress devices. Lastly, the authors have also considered how the internal synchronization of the 5G system influences the computation of the rate ratio. The simulation results highlight potential issues, especially the significant synchronization errors that can arise from even small frequency offsets between the ingress and egress points in the 5G network. The paper concludes by suggesting future work to address these challenges and improve time synchronization in integrated 5G-TSN networks.

\subsubsection{Latency and reliability}

In \cite{8643425}, an end-to-end 5G network architecture for manufacturing automation has been studied. The framework addresses key challenges, such as ensuring the 5G network meets stringent communication requirements of industrial control systems such as time synchronization and QoS, achieving interoperability with existing industrial protocols, and incorporating critical functions not yet supported by 5G. The paper details technical solutions, including latency and reliability improvements through Coordinated Multipoint (CoMP) transmission, protocol adaptations for TSN, and prototype validation using industrial Ethernet variants.
 In addition, a prototype system that uses the PROFINET Real-Time protocol to link genuine industrial control equipment (such as PLCs, motor controllers, and servo motors) is presented to verify specific elements of the design. The study demonstrates the criticality of round-trip latency for high-performance motion control systems. These findings underscore the necessity of millisecond-level round-trip latency for wireless communication systems to emulate the performance of Industrial Ethernet. Furthermore, time synchronization using PTP over a 5G link was found to be inadequate due to packet jitter, resulting in over 300 microseconds of inaccuracy, highlighting the need for over-the-air synchronization using native 5G radio to achieve the required 1 microsecond precision

In \cite{9779191},  a novel mechanism for merging TSN and 5G communication is presented with a focus on the translation of data flows among them. The goal is to translate incoming data frames between TSN and 5G networks using the user data plane and translation process. Additionally, a proof-of-concept model of the suggested method in the OMNET++ based TSN simulator NeSTiNg is presented. An automotive industry use case was developed to assess the effectiveness of the suggested approach and simulation. Based on the obtained results, the proposed mechanism may assist network designers in assessing the end-to-end latency of TSN-5G heterogeneous network setups.

In \cite{9828687}, a new approach for dynamic bandwidth management and TSN channel division in a 5G TSN network for a vehicular communication use case is presented. In this study, an edge-to-cloud intelligent vehicle-infrastructure cooperative system is created that enables minimal-latency, reliable multi-data source transfer and data integration between various cars as well as between cars and roadways. AI and digital twin technologies are also used in this approach. The technology uses artificial intelligence to establish twin auxiliary data and then uses digital twin representation to convey the findings to the vehicle in a straightforward and informative manner. To ensure the effectiveness of the system, it is evaluated using two application scenarios. To evaluate the performance of the system, three comparison tests are conducted: data transmission via 5G, 5G+TSN network transmission, and the proposed 5G+TSN transmission bandwidth dynamic calculation and TSN channel division method. These tests are performed in various scenarios to assess the system's architecture performance, including end-to-end service time, network delay, and system processing time. Other parameters, such as packet loss rate, response delay for blind area detection and intersection collision avoidance, are also evaluated. The findings indicate that the proposed method outperforms others in terms of overall performance.

In \cite{9065178}, a thorough overview of the integration of 5G and TSN is presented. Furthermore, to analyze the effects and determine the needs for the end-to-end system while handling time-sensitive deterministic traffic, a system-level simulator is presented that covers the primary 5G URLLC and TSN user plane functions. The simulator is then used to examine system performance in a multi-user instance. It was concluded that existing 5G mechanisms designed for URLLC traffic are insufficient to meet the demands of TSN streams. Consequently, an end-to-end scheduling approach encompassing both systems is required.  Authors in \cite{9779167} discussed the traffic steering issue in edge computing devices with 5G UEs in 5G-TSN networks. In this study, a new eXpress Data Path (XDP) programming use case to lower latency is provided, which is essential for achieving low E2E latencies. XDP functions as a high-efficiency data pathway designed for examining and directing packets along the downlink path. It leverages the extended Berkeley Packet Filter (eBPF) and has been integrated into the Linux kernel. The suggested use case for traffic steering latencies was put into practice on a 5G testbed and contrasted with an L2 bridge. The findings clearly show that XDP has the potential to improve performance by a factor of 100.

In \cite{10210522}, a novel deterministic Federated Learning (FL) framework known as DetFed is introduced,  tailored for time-sensitive industrial IoT environments leveraging 6G-oriented URLL communication techniques. 
The framework integrates a wireless and wired TSN infrastructure to ensure deterministic transmission of FL traffic while preserving the concurrent transmission of burst traffic, such as safety-critical data, which is crucial for the efficient aggregation and distribution of model parameters across multiple layers in the network. To address the challenges posed by the dynamic and resource-constrained nature of industrial IoT, the authors develop a deep reinforcement learning (DRL)-based dynamic resource scheduling algorithm, which optimizes temporal resource allocation, thereby accelerating FL convergence and enhancing the overall learning accuracy. This method is used to establish connections across three layers, and within each switch, a cyclic queuing and forwarding mechanism is implemented. This mechanism is designed to facilitate the deterministic transmission of model parameters with microsecond-level delays and almost zero packet loss, ensuring the reliability of the process. Comparing the proposed DetFed against state-of-the-art benchmarks, simulation results using real-world datasets demonstrate that it dramatically speeds up federated learning convergence and increases learning accuracy.

\subsubsection{Traffic scheduling and resource allocation}

In \cite{9138407}, a radio resource allocation technique for deterministic downlink time-sensitive communication streams using traffic pattern information is presented. To enable time-sensitive communications with dynamic packet scheduling, a measuring and reporting scheme based on the user equipment is utilized. Furthermore, a safety offset parameter to compensate for changes in the critical Signal-to-Interference-and-Noise-Ratio (SINR) that could potentially negatively affect TSN performance is added. In this study, a novel user equipment-based evaluation and reporting system is presented, in which evaluations are made on real data allocations throughout a time frame, for semi-persistent scheduling-enabled time-sensitive networks. The approach has been enhanced for Modulation and Coding Scheme (MCS) selection and utilizes the least SINR value which was recorded together with an offset at the 5G base station (gNB). A comparison of semi-persistent and dynamic packet scheduling techniques is provided, along with essential improvements to link adaptation and interference collaboration processes, considering physical layer control channel impacts in the assessment. Compared to utilizing conventional dynamic packet scheduling techniques suggested for URLLC traffic, this solution offers a gain of more than 3 times in terms of network capacity, allowing for a significantly higher number of supported users. Additionally, compared to a random distribution of users in each cell, the suggested soft frequency-reuse allocation mechanism across cells offers an additional gain of around 2 dB SINR. The proposed technique offers a three-fold benefit for a 20 bytes payload and two-fold gains for a 50 bytes payload in comparison to a specified fixed MCS reference scenario. In comparison to conventional dynamic scheduling approaches, which are often assumed for URLLC applications, the network capacity in terms of the number of supported time-sensitive streams may be nearly tripled using the suggested approaches.

In \cite{10062219}, a new Configured Grant (CG) scheduling strategy for 5G integration into a TSN, in which 5G serves as a bridge for the TSN, is proposed. The presented technique seeks to satisfy the various TSN flows' performance criteria such as determinism and end-to-end latency. In order to align its decision with TSN schedules, the suggested approach makes use of the information supplied by TSN on the characteristics of the TSN traffic. In \cite{10110348}, the authors introduced a novel approach for QoS mapping, utilizing an enhanced K-means clustering algorithm and the theory of rough sets. The method, referred to as Improved K-means Clustering based on Rough Set Theory for QoS Mapping (IKC-RQM), establishes a dynamic and load-aware QoS mapping algorithm to enhance its adaptability. Furthermore, an Adaptive Semi-Persistent Scheduling (ASPS) mechanism is implemented to address the intricate deterministic scheduling challenges within the context of the 5GS (5G System). This mechanism is divided into two main components: the first deals with allocating persistent resources for time-sensitive data flows, while the second handles dynamic resource allocation using the max-min fair share algorithm. Based on the simulations, the proposed algorithm effectively achieves versatile and suitable QoS mapping. Furthermore, the ASPS mechanism ensures appropriate resource allocation, guaranteeing the dependable transmission of time-sensitive data flows within integrated 5G-TSN networks.

In \cite{10136614}, the authors have proposed an innovative approach through the integration of TSN and 5G in an industrial network architecture. In this setup, the 5G system functions as a logical bridge with TSN capabilities. Building upon this network framework, a new algorithm called Double Q-learning-based hierarchical particle swarm optimization (DQHPSO) to tackle the optimization of scheduling is presented. The DQHPSO algorithm incorporates a population structure organized into levels and integrates Double Q-learning to modify the number of these levels within the population dynamically. This strategic adjustment effectively circumvents local optimization challenges, leading to an enhancement in search efficiency. Through extensive simulation experiments, the efficacy of the DQHPSO algorithm in increasing the success rate of scheduling for time-triggered flows is illustrated, surpassing the performance of alternative algorithms. In \cite{10275568}, a framework that combines essential elements to enable Network Digital Twins (NDT) in both public and private 6G networks utilizing TSN is introduced. The framework is designed to address the challenges associated with NDT deployment. Notably, the framework proposes a modular NDT system that can learn and adapt to create the digital twin model. It offers flexibility by leveraging existing open-source simulators with pre-built models, or it can directly learn the model from real-world network infrastructure using artificial intelligence (AI) and machine learning (ML) trained on collected data. Additionally, the paper emphasizes the significance of NDT, Software-Defined Networking (SDN), and deterministic networking approaches (TSN and DetNet) as cornerstones for building a robust future 6G network architecture with TSN features.

 In \cite{10637302}, a network function virtualization-enabled 5G-TSN framework is proposed to address the challenges posed by the disparate access mechanisms and scheduling granularities of 5G and TSN, thereby enabling unified resource management. In this framework, network flows are scheduled through network slicing and virtual network function embedding for centralized resource allocation. A novel metric, the full-path age of information (FP-AoI) model, is introduced to facilitate integrated scheduling within the 5G-TSN system by conceptually representing the 5G system as a sampling process of the virtual TSN network. To mitigate the issue of extended latency tails associated with 5G, the scheduling of 5G and TSN resources is formulated as a risk-aware FP-AoI minimization problem. To solve this NP-hard problem, a decomposition and augmentation-based joint scheduling (DAS) algorithm is proposed, which decomposes the problem into three sub-problems. The first two sub-problems are shown to be convex. For the third sub-problem, TSN scheduling, a FP-AoI-driven TSN scheduling scheme (FvQI) is designed by constructing an augmented logical topology that incorporates service function chain constraints and TSN characteristics. FvQI offers a computationally efficient TSN scheduling solution. The simulation results demonstrated improvements in reliability, runtime, and resource utilization, highlighting the efficacy of the proposed methods. However, the assumptions made for the simulation environment might not fully capture the complexities of deploying such a framework in diverse and dynamic network conditions, which could affect the applicability and scalability of the proposed approach.

\cite{10623103} focused on the joint time-frequency resource allocation problem in TSN-5G networks to ensure ultra-low latency and high reliability. The authors formulated the optimization problem as NP-hard and transformed it into a Markov decision process (MDP). A novel Dueling Double Deep Q-Network (D3QN) based joint resource allocation algorithm, called DJRA, was introduced to solve the problem and maximize scheduling efficiency while achieving throughput fairness. Through simulations, the DJRA algorithm showed improved convergence speed, computational efficiency, and overall performance compared to other benchmark algorithms. The paper’s approach, while innovative, could be limited by assumptions of ideal network conditions and may face challenges in real-world scenarios where dynamic environments and computational complexity impact performance. Additionally, issues like scalability, energy efficiency, and the lack of real-world validation or comparison with other state-of-the-art methods leave room for further exploration and improvement.

\subsection{Summary and lesson learned} \label{sec:SummarySec4}
Each wireless technology offers unique advantages and challenges for TSN implementation. For instance, Wi-Fi technologies such as IEEE~802.11ac and IEEE~802.11ax provide high data throughput and efficiency through features like MIMO, frame aggregation, and advanced modulation techniques. However, achieving precise time synchronization and deterministic behaviour in Wi-Fi networks requires innovative mechanisms such as low-overhead beacon-based synchronization and modification of the control software in Wi-Fi nodes. New OFDMA and multi-link features in IEEE~802.11be (Wi-Fi 7) will help with the implementation of TSN integration as it offers more flexible and adaptive resource allocation and enhanced reliability over wireless links. Similarly, 5G networks, with their support for URLLC and network slicing, offer promising avenues for TSN integration. Yet, realizing seamless integration between 5G and TSN demands robust synchronization mechanisms and traffic scheduling strategies, as evidenced by studies exploring 5G-TSN convergence. While some studies showed that bounded latency can be achieved, future research needs to consider a broader range of wireless conditions, e.g. interference, error rates, wireless channel conditions, to dynamically select robust wireless communication parameters to meet TSN reliability requirements.

Similarly, for Wi-Fi, the research addresses specific TSN requirements such as time synchronization, latency, reliability, and traffic scheduling within the context of each wireless technology. Solutions like low-overhead beacon-based synchronization mechanisms, TDMA-based protocols, and dynamic bandwidth management techniques demonstrate efforts to meet these requirements. Additionally, innovations in traffic scheduling algorithms, resource allocation strategies, and QoS mapping techniques are crucial for optimizing TSN performance over wireless networks. Newer version of Wi-Fi such as IEEE~802.11ax and IEEE~802.11be offer more enhanced radio resource management techniques that offer the potential for meeting TSN requirements. However, more research is needed to demonstrate that this can indeed be achieved for a variety of wireless environments. In particular, adaption to changing wireless conditions needs to be explored further. 

Furthermore, the studies highlight the significance of standardization efforts and collaborative frameworks like IEEE and 3GPP in shaping the development and deployment of wireless TSN solutions. These standards provide a common ground for interoperability and facilitate the integration of TSN capabilities into existing and future wireless technologies. Overall, the lesson learned emphasizes the need for a holistic approach that combines technological innovation, standardization efforts, and practical implementation strategies to realize the full potential of wireless TSN for time-sensitive applications across various industries.

\section{Wireless TSN Applications} \label{sec:App}
\begin{figure*}[t]
\centering
\includegraphics[width=5.25 in, height=2.5 in] {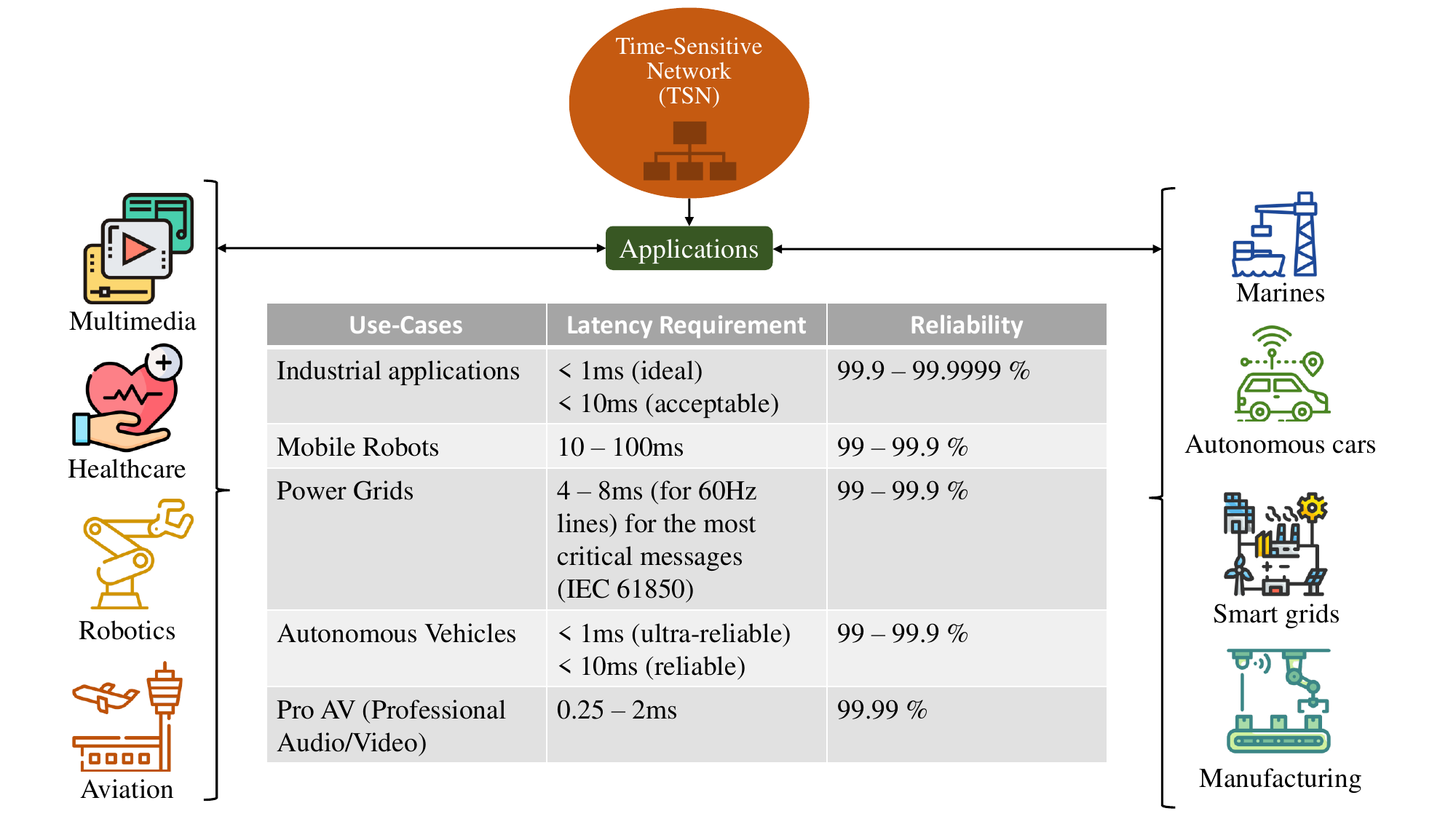}
\caption{Overview of wireless TSN applications}
\label{fig4}
\end{figure*}

The aim of wireless TSN is to offer ultra-reliable data transfer with deterministic and bounded end-to-end low latency for a broad range of applications. The current most common TSN application domains are shown in Fig.~\ref{fig4} and are described in the following subsections. In each subsection, we present existing works involving the respective application and associated use cases with a summary of references in Table \ref{Table:applications}.

\subsection{Industrial automation}
Industrial automation is seen as one of the key application areas for both, wired and wireless TSN networks.
In \cite{8672474}, a fresh viewpoint on expanding TSN functionalities across wireless networks for industrial automation is presented. The paper begins with examining industrial communication prerequisites and networks, incorporating a concise exploration of prevailing wireless standards prevalent in industrial settings today. The paper elaborates on existing and potential extensions of TSN capabilities over wireless networks, encompassing aspects such as time synchronization, time-aware scheduling, reliability features, and the necessary adaptations for effective operation in the wireless domain. Additionally, it offers a framework for integrating wired and wireless TSN features into upcoming industrial automation systems as well as a taxonomy of wireless applications. Authors in \cite{en14154497} introduced two potential wireless TSN designs for 5GS and WLAN integration into industrial automation. In the first scenario, a WLAN interface is utilized to connect to the TSN node, while in the second scenario, a WLAN is connected to a 5GS in addition to the 5G RAN. Additionally, the technical challenges associated with wireless TSN for each of the considered standards are investigated. In \cite{9409838}, the primary focus of the paper's discussion is on Release-16 of the standard that relates to low-latency and high-reliability capabilities enabling industrial automation. The TSN and time synchronization accuracy constraints are also pointed out and the effectiveness of the 5G NR and data traffic streams with regard to ultra-reliable and low latency communication requirements are then evaluated. In \cite{10035542}, a network scheduling strategy for the convergence of TSN and industrial wireless networks is presented. The approach consists of an SDN-based convergent network framework for converting data flow priorities and a technique for scheduling the corresponding data flows in industrial wireless networks and TSNs. Authors in \cite{8757992} present a mechanism to modify the MAC layer to improve Wi-Fi's Quality of Service for the downlink while simultaneously monitoring the network's jitter and delay performance. According to the simulation results, adding a new access category for time-sensitive traffic enhances performance and makes it possible to use TSN and Wi-Fi technologies in a variety of industrial automation use cases.

In \cite{9483597} the authors have explored the deployment of a wireless-wired TSN network called SHARP. It focuses on enabling hard real-time capabilities with bounded latencies in the range of hundreds of microseconds and achieving clock synchronization accuracy down to tens of nanoseconds. These capabilities are applicable to both wired and wireless segments, facilitating the extensive digitization of highly adaptable smart factories in the future. In \cite{9921680}, a 5G and TSN-enabled industrial control system is described and evaluated. The paper outlined an approach with TSN over a 5G test bed that was created with 3GPP Release 15 compliant hardware, and the measurement technique and key performance indicators necessary to satisfy the timing requirements for an industrial control application were gone through. The techniques and TSN over 5G system configurations required to define the Quality of Experience (QoE) for commercial use cases were also examined. They demonstrated that enabling TSN features over 5G, such as IEEE~802.1AS time synchronization and IEEE~802.1Qbv scheduling, significantly improves the latency and reliability of time-sensitive industrial applications, even under high network load conditions. In \cite{9921559}, the combination of PROFINET with wireless TSN is described. By delivering deterministic communication and dependability under best-effort traffic congestion in the wireless network, the support for safety-concern use cases that can be provided without any issues is demonstrated. The approach is assessed based on the obtained end-to-end latency and the reliable communication's hourly failure rate. Based on the results, the utilization of wireless TSN, incorporating dedicated time slots for each traffic flow, can lead to the attainment of safety integrity levels reaching grade 4. In \cite{9297066}, an overall framework to allow the design of TSN in wireless industrial networks with an emphasis on the fully centralized architecture is discussed. For this purpose, a configuration entity is built to configure the wireless end nodes in accordance with their needs. The Digital Enhanced Cordless Telecommunication (DECT) ultra-low energy (ULE) protocol is then used to verify the suggested solution. 

In \cite{10144121}, a dynamic traffic classification mechanism to provide quicker exclusive access to the wireless medium for packets of extremely time-sensitive flows, which may be produced arbitrarily is proposed. To give particular packets priority over others when it comes to channel access, dynamic traffic classification makes use of a mechanism known as shadow queues that are utilized in the openwifi FPGA-based Wi-Fi baseband SDR platform. In the study, it was demonstrated that the latency for accessing the channel remains unaffected by the scheduling cycle when considering dynamic traffic classification. Instead, it relies on the arrangement of dedicated time slots within the scheduling cycle. Consequently, the overall latency is reduced in scenarios involving longer communication cycles and greater intervals between communication time slots. 
In \cite{app13137953}, a comprehensive three-layer communication architecture for virtual power plants, integrating 5G and time-sensitive networking (TSN) to ensure both determinism and mobility in communication, is introduced. The paper has provided an in-depth analysis of service types and traffic requirements within the virtual power plant context, establishing a seamless mapping between 5G and TSN networks to uphold the QoS. In addition, a novel approach featuring semi-persistent scheduling with reserved bandwidth sharing and a pre-emption mechanism to cater to time-critical traffic, guaranteeing bounded latency and enhanced bandwidth utilization, is proposed. The performance assessment results have demonstrated the effectiveness of this mechanism in reducing end-to-end delays for both time-triggered and event-triggered traffic compared to dynamic scheduling methods.

\subsection{Robotics}
In the following, we present wireless TSN research focused on applications in robotics. The challenges in ensuring accurate time synchronization and deterministic data delivery over wireless channels are addressed in \cite{9473897}, using collaborative robotics as an illustrative application. In the paper, a mechanism is described for mapping QoS requirements at the application layer from the ROS2 (Robotics Operating System 2) and DDS (Data Distribution System) middleware, which are commonly employed in robotics application development, to the link layer. The link layer transport is based on wireless TSN capabilities implemented based on Wi-Fi. Experimental results from a prototype implementation of a collaborative task involving two robots, facilitated by wireless TSN time synchronization and traffic shaping over Wi-Fi, are provided in the paper. The findings indicated that even in situations where background traffic uses the same channel, time-synchronized and time-aware scheduling over Wi-Fi can be configured to partially satisfy the QoS needs of the robotics application. Despite significant advancements, several unresolved issues remain. Achieving precise QoS mapping between ROS2/DDS and other wireless technologies like 5G is still a challenge, requiring further research. Additionally, ensuring efficient, scalable QoS in large-scale systems with multiple collaborative robotic clusters is complex and not yet fully addressed. Lastly, the integration of diverse systems to achieve end-to-end determinism in more complex and heterogeneous environments remains an ongoing area of exploration. In \cite{9483447} and \cite{9714203}, using wireless TSN technology, a collaborative robotic work cell test bed is presented and the difficulties associated with deployment, performance assessment, and management recommendations are discussed. The paper describes the procedures for putting major wireless TSN features such as time synchronization and time-aware scheduling into practice and evaluating their effectiveness over IEEE~802.11. The wireless TSN capabilities are implemented on the collaborative robotic work cell test bed at the National Institute of Standards and Technology (NIST), which consists of a pair of robotic arms and simulates an object-handling application referred to as machine tending. The study focused on examining various configurations and measurement methods to evaluate the QoE associated with this use case. Furthermore, it aimed to establish a correlation between the QoE and the wireless network's performance. In \cite{10144124}, utilization of the TSN redundancy feature as specified in the IEEE~802.1CB, to remove failures or delays caused by occurrences like roaming and interference in a robot with mobility use case supported by Wi-Fi 6 TSN, is demonstrated. Through simulation evaluations and experimental results with a mobile robot connected through multiple Wi-Fi 6 transmitters in an industrial setting, the results demonstrated that while roaming performance was improved, challenges remain in ensuring that application performance is not negatively affected, particularly under real-world conditions and with commercial hardware implementations.

In \cite{electronics11111666}, a 5G system design and deployment with TSN, along with the investigation on a common industrial application such as cloud-controlled mobile robots, is presented. In order to analyze the effectiveness of 5G-TSN for industrial use cases, a prototype configuration incorporating 5G in a TSN network has been constructed. The architecture is covered in a comprehensive manner, along with a detailed explanation of the jitter reduction techniques used in the communication chain involving the mobile robot and the cloud-hosted application used in a manufacturing facility. With over-the-air testing on a commercial shop floor employing TSN features like traffic shaping and scheduling, the combined 5G and TSN prototype has been assessed. The study demonstrates that the integration of 5G with TSN networks in industrial environments effectively reduces latency and jitter, achieving the high reliability and low-latency communication required for mobile robotics. The experimental validation showed that the integrated system could maintain end-to-end latency below 0.8 ms with high precision in time synchronization. However, the findings also revealed remaining challenges, particularly the significant jitter introduced by the 5G system during over-the-air transmissions, which reached up to 500 \(\mu\)s.
 In \cite{9968494}, a scenario is introduced for a cooperative robot heavy lift procedure in which two robots work together to move a heavy object in three dimensions using a TSN-capable IEEE~802.11 WLAN. Within the robot operating system (ROS) as the software middleware, scheduling is carried out via IEEE~802.1Qbv over WLAN. To support the time-sensitive traffic of the robotic application and to enable high data rate traffic from IIoT applications to coexist, the technique for schedule selection is outlined in this study. Furthermore, the joint robot application's operational effectiveness which is impacted by the TSN scheduling selection is evaluated. The findings of this study demonstrate that while the application of wireless TSN capabilities in an 802.11 network can effectively support a collaborative robotic use case in industrial environments, several challenges remain. Notably, the implementation of the 802.1Qbv scheduling feature requires meticulous planning and configuration of protected window sizes to account for various overheads, including channel access and transmission delays. Additionally, the use of TCP transport presents further challenges, as it necessitates consideration of TCP acknowledgments in the network schedule, leading to increased overhead. In \cite{10144232}, an efficient TSN scheduling mechanism with a focus on bandwidth is presented in order to handle data transmissions between two synchronized robots carrying an item together. In the study, using an IEEE~802.11 wireless medium, IEEE~802.1Qbv TSN scheduling is demonstrated, which ensured that robot performance requirements are met despite supporting simultaneous best-effort traffic streams. In addition to discussing the procedure for choosing a schedule and gathering experimental data, this article also discussed how to customize and optimize the parameters of a TSN setup. The findings of this study highlight the effectiveness of 802.1Qbv scheduling in accommodating TCP acknowledgments and efficiently utilizing channel bandwidth in wireless time-critical industrial applications. However, several challenges remain. The results indicate a need for further optimization of the configurations to achieve tighter control over the best-effort traffic window, preventing packet overflow into the protected time-critical window. The approach also demands stringent control of channel conditions to maintain the necessary Modulation and Coding Scheme (MCS) levels; under extreme conditions, the time-aware schedule must rapidly adapt to ensure reliable communication.

\begin{table*}[htbp]
\caption{Summary of existing work based on their use-cases and applications}  \label{Table:applications}    
\scriptsize
\centering          
\begin{tabular}{|c|c|}  
\hline 
\textbf{Application} & \textbf{Use-case} \\[0.5ex]  
\hline \hline
\textbf{Industrial} & \cite{8672474}, \cite{9065178},\cite{en14154497}, \cite{9409838}, \cite{10035542}, \cite{8757992}, \cite{9483597}, \cite{9921680}, \cite{9921559}, \cite{app13137953}\\ \hline
\textbf{Robotics} & \cite{9473897}, \cite{9483447}, \cite{9714203}, \cite{10144124}, \cite{electronics11111666}, \cite{9968494}, \cite{10144232}\\ \hline
\textbf{AGV} & \cite{9644616}, \cite{10012491}, \cite{electronics11030412}, \cite{10060159}, \cite{Bao2022} \\ \hline
\textbf{AR/VR} & \cite{Kim2018}, \cite{10620809}\\ \hline

\end{tabular}
\end{table*}

\subsection{Automated guided vehicles (AGVs)}
While in an industrial setting, a variety of use cases with different communication needs, such as process monitoring, video inspection, and tracking of objects, may coexist, stringent performance requirements for mobile equipment such as drones or automated guided vehicles (AGVs) need to also be met, which complicates network design and administration. In \cite{9644616}, the prospect of combining TSN wired networks with unaltered, best-effort wireless networks for Vehicle to Everything (V2X) communication is explored, which motivated a methodology that pushes complexity to the network edge. In this study, the outcomes of a straightforward proof-of-concept system simulation, motivated by a platooning use case are shown. To demonstrate the suggested idea, the focus is directed towards the scenario of vehicular platooning. In this case, a group of autonomous vehicles cooperatively forms a convoy and shares sensor data and manoeuvre information on a closely coordinated schedule. This enables the entire platoon to function as a single entity, allowing them to drive with minimal gaps between vehicles. Moreover, the challenges presented by unsynchronized In-Vehicle Networks (IVNs) are analyzed and the advantages of synchronized In-Vehicle Networks schedules compared to a standard IEEE~802.11p connection are illustrated. Additionally, the initial findings from a simulation-based investigation using the widely used simulators OMNeT++ with INET, Plexe, and Veins, are presented. The proof-of-concept simulation focuses on vehicular platooning, demonstrating that synchronizing IVNs across vehicles can significantly reduce latency in data transmission, from 100 ms to less than 1 ms. The study suggests that moving complexity to the network edge, rather than overhauling core and access networks, is a more practical approach. In \cite{10012491}, a TSN Controller (TSNC) NetApp mechanism is presented that can handle various vertical applications while guaranteeing QoS throughout their lifespan. The NetApp consisted of two entities: a controller unit that receives and handles requests from vertical applications with particular network performance requirements, and an agent unit that utilizes settings and keeps track of the status of network components. This control framework has been expanded to accommodate openwifi-based wireless TSN communication, enabling the flexibility needed by portable nodes used in vertical applications, such as AGVs and drones. The evaluation results demonstrated the feasibility of a Controller NetApp and assessed its resource usage, revealing its capacity to effectively manage networks consisting of lots of nodes while minimizing resource consumption. Additionally, the TSN extensions of openwifi are investigated and the evidence of the TSNC's capability to intricately manage traffic behaviour for wireless clients is presented. 

In \cite{electronics11030412}, the implementation of 5G is evaluated using actual industrial environment scenarios and deployment configurations. A 5G prototype system with pre-commercial and standard-compliant URLLC capability, as well as implementations utilizing commercial 5G systems, have been studied. Systematic evaluations were conducted to assess the performance of 5G in terms of latency and reliability. These evaluations encompassed various scenarios, including various packet sizes, devices, and networks with varying capabilities across multiple sites. The measurements were carried out both over the air and in live networks to obtain a comprehensive understanding of the expected performance. Additionally, a specific set of experiments was performed to empirically analyze the advantages of incorporating Frame Replication and Elimination for Redundancy (FRER), an element of the TSN standards. As part of the 5G-SMART project, two distinct 5G use cases implemented at the Bosch semiconductor factory in Reutlingen, Germany, are assessed in this study. The first use-case centres around cloud-based mobile robotics, specifically exploring the feasibility, flexibility, and performance of wirelessly controlled AGVs connected to the 5G network. The aim was to assess the potential advantages and capabilities of utilizing 5G connectivity for efficient AGV operations in an industrial setting. The research findings suggest that 5G systems, equipped with enhancements such as higher subcarrier spacing, optimized TDD patterns, advanced scheduling, prioritization, robust control and data channels, and faster control feedback mechanisms, may be capable of meeting the stringent latency requirements of demanding industrial use cases. However, early 5G networks and devices primarily designed for MBB services may experience increased latency when higher reliability guarantees are sought. The achievable latency is influenced by both the 5G device and the network configuration. Additionally, redundant transmissions, such as those enabled by FRER, can effectively reduce end-to-end latency in scenarios where high reliability is paramount. 

In \cite{10060159}, the clear preference for incorporating 5G into TSN and its various applications, including swarm robots and AGVs, is highlighted. Additionally, the 3GPP standard 5G-TSN specifications are assessed. An analysis of the research conducted on the topic, encompassing simulation tools, open-source 5G systems, and the existing hardware options accessible for integrating these elements, is provided. Based on the findings, the results reveal that existing research heavily relies on simulation outcomes or the possibility of establishing a strong collaboration with a limited number of hardware vendors. This reliance is attributed to the scarcity of commercially available hardware that can be used for evaluating the performance of 5G-TSN integration. In \cite{Bao2022}, the shortcomings of existing vehicle networks are examined and the state-of-the-art in TSN is discussed. In this paper, a TSN-based backbone network of train control management system (TCMS) is presented. Future autonomous vehicles may accomplish high-quality data transfers by integrating TSN into the TCMS backbone network. To prevent the control data from being impacted by the multimedia data, in this study, the priority of the vehicle data is also assessed and the time-aware shaper slots are determined. Based on the findings from experiments, the proposed framework may increase the network efficiency in rail transportation systems.

\subsection{Augmented reality (AR) and Virtual Reality (VR)}
Technology has advanced to allow users to fully immerse themselves in their virtual surroundings thanks to advancements in virtual reality technology, offering experiences that transcend the limitations of the physical world by creating entirely new environments and sensations. In the context of AR/VR, there are numerous potential advantages to utilizing TSN, although which applications will benefit most and how is still an open question \cite{10012491}. Among the conceivable applications of AR/VR that will benefit from TSN are those that aid industrial maintenance and training \cite{10275568}. TSN, in this context, plays a key role in ensuring that AR-related data, including overlays and instructions, are transmitted with minimal delay, thereby amplifying the efficacy of training and maintenance duties. Additional uses of AR/VR with TSN include the healthcare sector, such as remote surgery and illness diagnostics \cite{Torres_Vega2020-xg}.

In \cite{Kim2018}, an innovative virtual reality-based cyber-physical education system called TSRTnet, for effective learning in a virtual environment on a mobile platform is presented. Utilizing digital twin technology, in particular, the proposed mechanism may connect the actual world towards virtual reality. In the virtual reality service layer, the main service needs in terms of delay time are identified. In addition, a novel, real-time network technology interworking software, referred to as network and TSN, with the goal of fulfilling the demands of the network layer is proposed. To provide protocol-level accessibility, a gateway mechanism for interworking is proposed. A route selection technique is also suggested to provide flexible connectivity between real-world objects and virtual objects. In order to verify the effectiveness and efficiency in terms of packet loss and latency, a simulation study is performed. TSRTnet exhibited superior network performance compared to conventional best-effort networks. Specifically, it demonstrated a significantly lower variance in end-to-end delay, with fluctuations confined to within 2 milliseconds. Moreover, TSRTnet experienced a markedly lower rate of packet loss, thereby enhancing its overall reliability and suitability for applications demanding low latency and high data integrity. While the paper proposed an innovative framework but lacks discussion on potential challenges, particularly regarding the scalability and real-time performance of TSRTnet in complex, real-world scenarios, which could affect practical implementation.

In \cite{10620809}, the authors developed a testbed that integrates 5G-TSN connectivity with haptic input devices to simulate a remote surgery scenario, highlighting the potential of 5G-TSN in applications requiring ultra-low latency, high reliability, and precise control. The demonstration allows users to interact with a virtualized dental surgery environment using a 3D haptic pen, providing real-time feedback on latency and jitter changes. This setup showcases the ability of 5G-TSN to support demanding Tactile Internet applications by minimizing network jitter and ensuring bounded latency through optimized TSN configurations. The findings emphasize the transformative potential of 5G-TSN in fields such as telemedicine, where real-time tactile feedback and low-latency communication are critical. However, while the demonstration effectively illustrates the capabilities of 5G-TSN, it is conducted in a controlled environment, and the practical challenges of deploying such systems in real-world scenarios with varying network conditions are not fully addressed.

\subsection{Summary and lesson learned} \label{sec:SummarySec5}
The critical lessons learned from exploring wireless TSN applications, particularly in domains such as industrial automation, robotics, and autonomous vehicles, is the importance of seamless integration and adaptation of TSN capabilities across diverse network environments. The research highlights the necessity of considering each application domain's unique requirements and challenges when extending TSN functionalities to wireless networks. This means, the features of wireless TSN must meet a broad range of application requirements, ranging from ensuring ultra-reliable data transfer to meeting stringent latency constraints. This requires new approaches to integrated network management between 5G and TSN and also TSN and other wireless technologies. Novel management algorithms are required that can adapt the network parameters and network operation to changing applications demands while the network is in operation. Currently, TSN designs tend to be static, but with mobile TSN end devices and dynamic applications in a wireless TSN environment, novel management algorithms must be developed to cope with these new challenges for traditional TSN networks.

Furthermore, the studies underscore the significance of practical implementations and performance evaluations in real-world scenarios, shedding light on the feasibility and effectiveness of TSN in addressing the specific needs of industrial automation, collaborative robotics, automated guided vehicles, and augmented/virtual reality applications. Overall, the journey of wireless TSN elucidates the importance of tailored solutions and continuous advancements to harness the full potential of deterministic communication in diverse wireless environments, paving the way for enhanced efficiency, reliability, and innovation across various industries.

\section{Open research challenges and future directions} \label{sec:Open}


Although wireless TSN promises real-time deterministic communication with very low latency, ultra-reliability, flexibility, and mobility, several technical challenges need to be tackled for wireless TSN to fulfil its promise. When these have been addressed, we expect the wider adoption of wireless TSN networks in various application domains such as smart manufacturing and automation, vehicular networks and AGVs, robotics, VR/AR/XR, etc. In this section, we discuss major challenges underlying the extension of TSN capabilities over the wireless domain informed by this literature review and propose future research directions to address them.

\subsection{Wireless nature of the environment}

A major challenge in extending TSN features into the wireless domain arises from the wireless nature of the communication channel. A wireless communication channel exhibits several impairments in terms of interference, signal disturbance and attenuation, multi-path fading, etc \cite{beard2016wireless}. Furthermore, the shared nature of the wireless medium with the typical coexistence of multiple wireless systems/devices operating in the same frequency band can lead to increased interference and signal disturbance beyond natural causes, resulting in increased packet losses. Operation in unlicensed bands usually results in a higher level of interference than in licensed bands due to the frequent coexistence of multiple systems and technologies. Hence TSN over Wi-Fi will likely have to deal with increased interference compared to TSN over 5G or future 6G cellular. 

The operating environment and its dynamics also pose several challenges for wireless communication, such as physical obstructions, metallic surfaces in industrial environments, etc., resulting in unwanted reflections, diffraction, and scattering, leading to multipath fading, blockage of wireless signals, reduced coverage, fluctuations in the link quality etc. The approaches/schemes used to attain various TSN capabilities, such as time synchronization, bounded end-to-end latency and ultra-reliable communication, need to take these challenges into account. Hence, future research needs to consider designing approaches and algorithms that can achieve TSN capabilities irrespective of the dynamic interference/disturbance caused by the wireless medium. As discussed in section \ref{sec:Tech}, different wireless technologies have various features enabling TSN capabilities over the wireless domain. For example, Wi-Fi 7 features multi-link operation, multi-AP coordination, enhanced OFDMA, etc., whereas network slicing, flexible numerology, and guaranteed QoS support are some of the key enablers of TSN over 5G. Furthermore, the characteristics specific to the particular environment/application need to be considered and accommodated in the design and implementation of a wireless TSN network. All this requires careful study of how best to implement TSN features into the different wireless technologies as this may not be the same across the technology spectrum.

\subsection{Channel access \& resource allocation}

Guaranteed channel access and deterministic scheduling of wireless resources are essential elements to achieving TSN capabilities over wireless communication channels. Traditional contention-based medium access control schemes, as have been used in Wi-Fi, are not able to facilitate deterministic latencies to the level of sub-milliseconds that certain time-critical TSN flows demand. Several new features, such as MU-MIMO, OFDMA, MU EDCA Parameter set, Target Wake Time (TWT) and restricted-TWT, Triggered TXOP sharing, etc., have been introduced in recent Wi-Fi standards (IEEE~802.11ax and IEEE~802.11be) that can help to reduce the probability of collisions among multiple stations in Wi-Fi. However, there can be specific scenarios where a series of unfortunate events (back-to-back transmissions by other stations) can prevent a station from transmitting its time-sensitive packet at the right time, as detailed in \cite{10070404}. Standards usually do not specify implementation approaches, but these often play a pivotal role in achieving the standards objectives, such as guaranteed channel access. Hence, we believe there is still a need for further study into guaranteed channel access mechanisms/approaches to enhance the existing protocols standardised by the IEEE for Wi-Fi.

Efficient wireless resource management is also crucial in extending TSN over the wireless domain. Wireless resources need to be allocated optimally so that all  TSN flows meet the required QoS metrics in terms of latency and reliability. The QoS definition itself is not well aligned across different wired and wireless technologies, and a standardised mapping needs to be established. The resource allocation algorithms also need to account for the variations in the wireless channel and the potential mobility of devices. Although the method of channel scheduling varies based on the wireless technology employed, a common need is to manage the timing of channel occupancy, which can be integrated with the time slot allocation conducted by TSN at the service level. As mobile devices require seamlessly connected handover between access points/base stations, the association/ disassociation process must also be fast. 
Optimal wireless resource allocation algorithms that can address these challenges are open research areas for TSN over 5G and Wi-Fi 7. 

Although TSN standards mention three different network management strategies (centralised, distributed and hybrid), most of the existing research follows a centralised approach. In the centralised management approach, different network elements in the TSN network, such as listeners/talkers and network bridges/switches send their capabilities and TSN flow requirements to the central management entity (CNC) where the network configuration/management decisions are made. This approach may not work well when end devices are highly mobile, getting connected to different access points at different times, requiring frequent network reconfiguration. Thus, in addition to centralised schemes, both distributed and hybrid network management schemes must be studied to identify how best to handle mobile TSN network scenarios. This also plays a vital role as a network's scale increases; another topic for further research in this area. 

\subsection{Interoperability \& integration with other wireless networks}

Many of the application domains discussed in Section \ref{sec:App} exhibit coexistence of several communication technologies (both wired and wireless). For example, a smart manufacturing application will have many devices on the shop floor, operating with different wireless technologies such as Wi-Fi (different variants), private 5G, Zigbee, Bluetooth, WirelessHART, etc., requiring them to coexist or interoperate \cite{s23010073}. Many of them may, in fact, be part of a wireless TSN network. Integrating a wireless TSN network with other existing industrial networks, including legacy systems/devices, and handling their interoperability aspects without negatively affecting various TSN features (discussed in Section \ref{sec:background}) is a wide-open topic that offers many opportunities for future research.

The interoperability challenge calls for new approaches or interfaces that can provide better compatibility between coexisting heterogeneous network protocols. Features such as the 5G LAN-type service \cite{10225780} and middleware platforms play a vital role in handling the interoperability between various systems. Integrating TSN into such middleware platforms is an important area that needs to be studied. Software-defined networking and network virtualization are two major trends in wireless networks, where different network functionalities can be hosted on on-premise generic hardware or in the cloud and easily reconfigured. How well TSN capabilities can be implemented using SDN/network virtualization in a standardized manner, meeting all the QoS requirements of various TSN traffic flows has yet to be addressed and is another future research direction.

\subsection{Traffic classes \& schedule management}

TSN aims to handle time-critical real-time traffic that demands very low latency and ultra-high reliability along with best-effort traffic through various mechanisms discussed in Section \ref{sec:Tech}. This is mainly achieved by assigning different priorities to traffic flows based on their QoS requirements and by scheduling the opening/closing of the ingress and egress ports of various network elements/switches in the communication path in a precise time-synchronized fashion. When the network is planned and implemented carefully, everything should go smoothly as long as there are no changes in the predefined network topology or the traffic classes. When a new type of traffic is introduced, or the network topology changes at runtime (for example, a new type of device is added to the network), which is common in wireless networks, mechanisms for flexibly integrating these changes into the TSN network at runtime have yet to be developed. A new TSN configuration schedule may have to be generated without affecting the ongoing time-critical message exchanges, which is another important research challenge.

In addition to TSN configuration schedule generation, routing schemes that span multiple TSN networks providing deterministic latency and ultra-reliability are major research areas in wireless TSN. Most of the wireless TSN research reported in the literature concentrates on the time synchronization aspects. In addition to time synchronization, future research must address the routing, scheduling and reliability aspects of wireless TSN. In addition to addressing each of these aspects individually, joint approaches also need to be studied. How well a proposed schemes can adapt to variations in the network size and topology, traffic flow requirements, and the dynamics in the application environment are other important aspects that require further study.




\section{Conclusion}\label{sec:Con}
This paper has offered a comprehensive survey and evaluation of the existing literature on the extension of Time-Sensitive Networking (TSN) for wireless communication systems (e.g. Wi-Fi and 5G) and the deployment of wireless TSN in various application domains. While an in-depth examination of the existing literature highlights continuous research efforts, noticeable research gaps remain in tackling fundamental issues associated with deploying wireless TSN in real-world scenarios. The latest Wi-Fi version, Wi-Fi 7 (IEEE~802.11be) is expected to be accompanied by a clearly defined and backwards-compatible time-sensitive operation mode to facilitate low-latency communication. Given the utilization of license-exempt frequencies, Wi-Fi will likely never be able to guarantee 100\% deterministic communications, but there are ways to minimize latency through optimal control of protocol operations. Additionally, the integration of 5G with TSN, while promising for enhancing real-time communication in industrial networks, also presents challenges such as time synchronization, resource management, and flow management, as highlighted in the recent literature. These challenges underscore the need for continued research, particularly in developing prototypes and practical tools that effectively implement 5G-TSN integration, as well as in evaluating these approaches in both simulated and real network environments. Resources must be distributed wisely to ensure that all TSN flows adhere to the necessary QoS criteria for latency and reliability. The inherent fluctuations in a wireless channel and the potential device mobility in wireless networks must be taken into consideration when designing resource allocation mechanisms. The interaction and collaboration between wireless scheduling, resource management, and the overarching scheduling of the entire TSN network all remain open issues that necessitate further exploration.


\bibliography{references}
\bibliographystyle{IEEEtran}

\begin{IEEEbiography}
[{\includegraphics[width=1in,height=1.75in, clip, keepaspectratio]{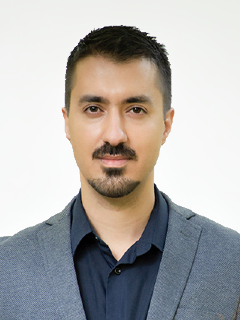}}]{Kouros Zanbouri} received a BSc degree in Information Technology (IT) engineering in 2016 and pursued an MSc degree in IT engineering - specializing in Computer Networks, in 2018. Kouros is currently conducting research on Time-Sensitive Networking (TSN) as a Ph.D. Student at UCC, in CONNECT Centre, School of Computer Science \& IT, University College Cork, Ireland.
\end{IEEEbiography}
\vskip 0pt plus -1fil 
\begin{IEEEbiography}[{\includegraphics[width=1in,height=1.75in,clip,keepaspectratio]{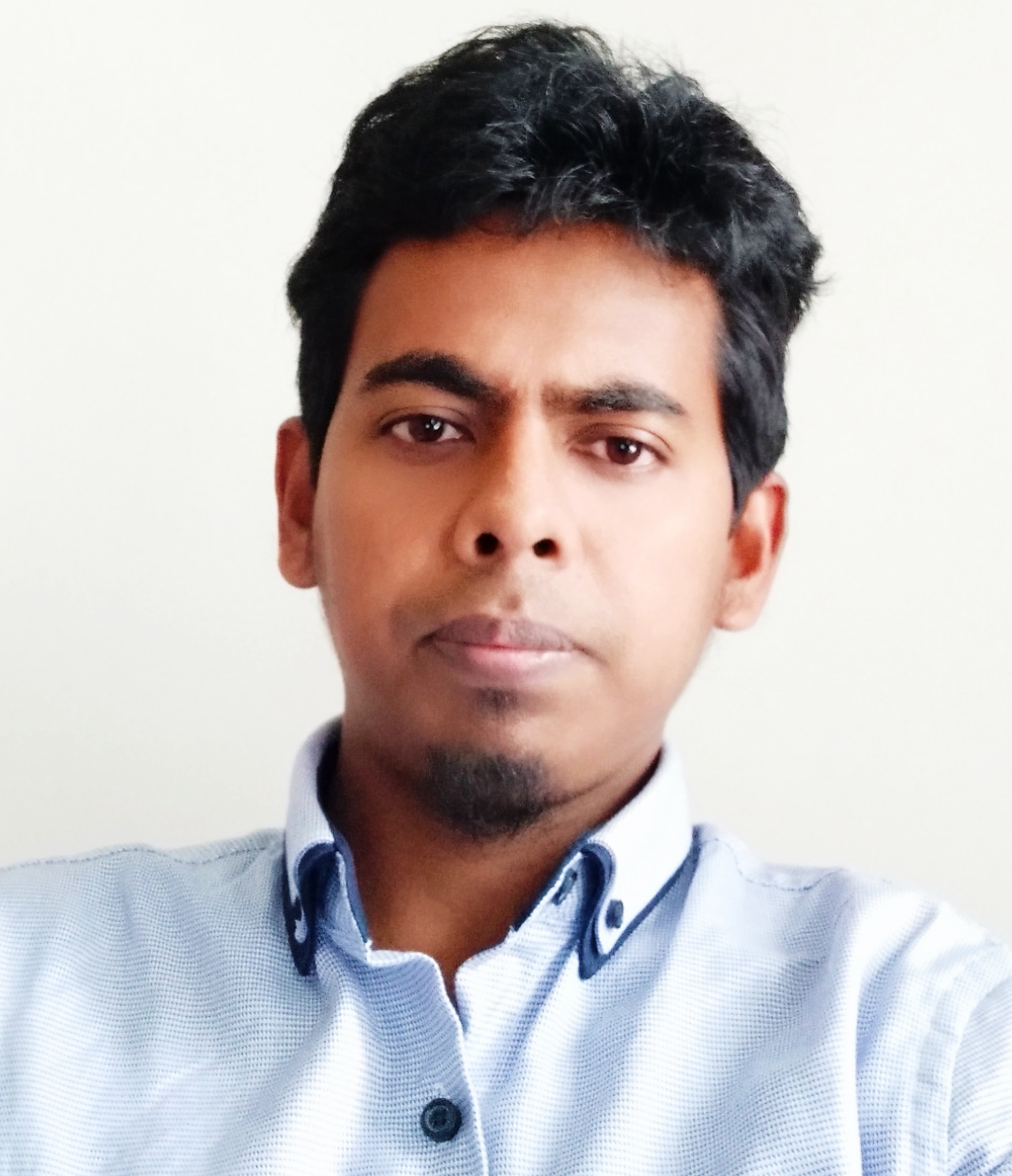}}]{Md. Noor-A-Rahim} received the Ph.D. degree from the School of Information Technology and Mathematical Sciences, University of South Australia, Australia in 2015. Dr. Rahim is currently serving as an Assistant Professor (Lecturer-Above the Bar) within the School of Computer Science and Information Technology at University College Cork, Ireland. Prior to this role, he was a Senior Researcher and Marie Curie Fellow within the same school. He also held the position of Postdoctoral Research Fellow at the Nanyang Technological University (NTU), Singapore. His research interests include Control over Wireless Networks, Intelligent Transportation Systems,  Machine Learning, Signal Processing, and DNA-based Data Storage. In recognition of his academic excellence, he was honored with the Michael Miller Medal for presenting the most outstanding Ph.D. thesis in 2015. For more details, please visit: https://narahim.github.io/
\end{IEEEbiography}
\vskip 0pt plus -1fil 
\begin{IEEEbiography}[{\includegraphics[width=1in,height=1.75in, clip, keepaspectratio]{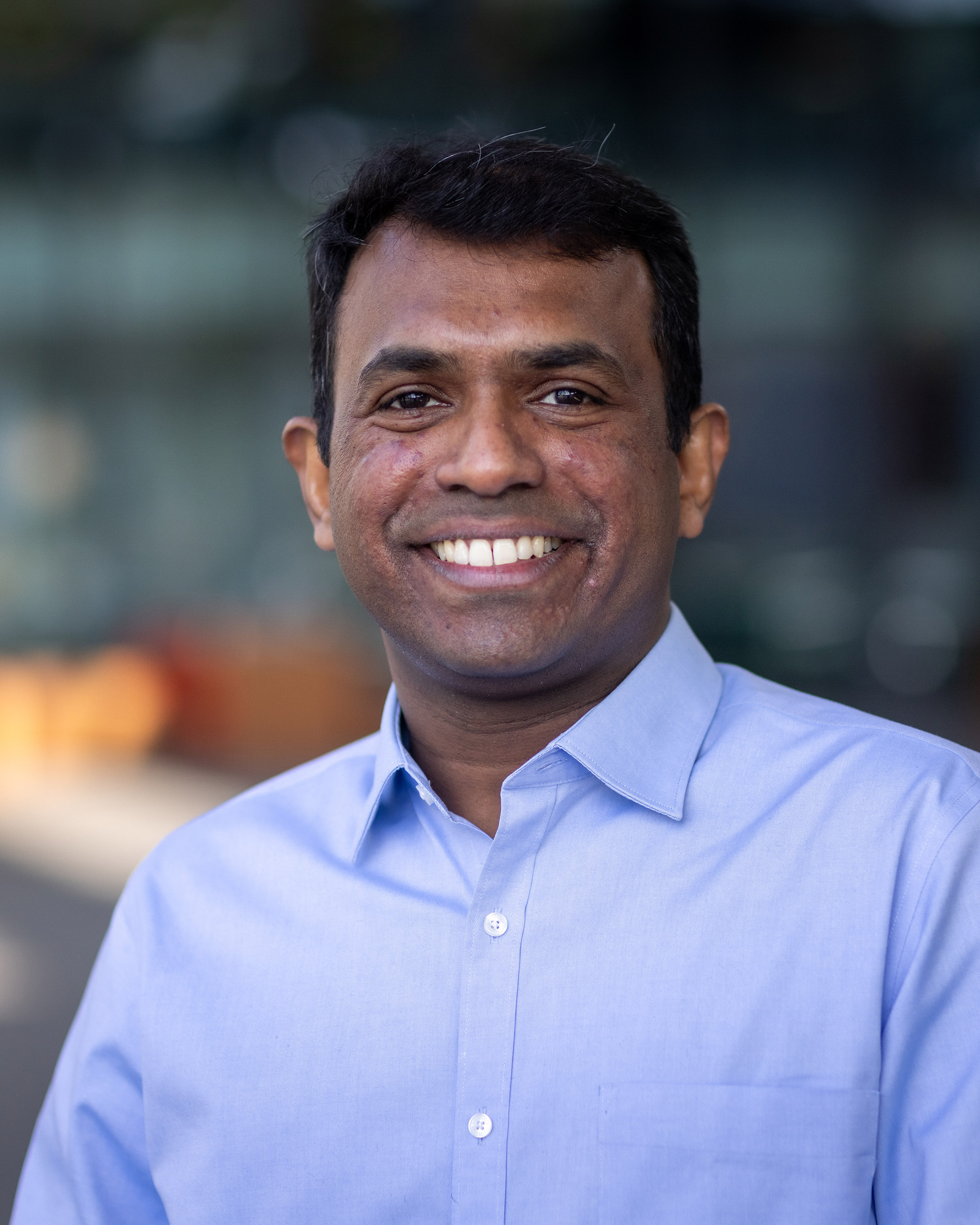}}]{Jobish John} received Ph.D. in Electrical Engineering from Indian Institute of
Technology, Bombay in 2020. He is currently working as a post-doctoral researcher at The Centre for Wireless Technology Eindhoven (CWTe), Department of Electrical Engineering, Eindhoven University of technology, The Netherlands. Before this role, he was a Marie Curie Fellow with the School of Computer Science and Information Technology, University College Cork, Ireland. His main research interests lie in the area of next-generation wireless networks. This ranges from ultra reliable and low latency communication systems for various verticals such as avionics, Industry 4.0/5.0, to Internet of Things and Low -Power Wide Area Networks. to distributed, heterogeneous, self-organizing networks.
\end{IEEEbiography}
\vskip 0pt plus -1fil 
\begin{IEEEbiography}[{\includegraphics[width=1in,height=1.75in, clip, keepaspectratio]{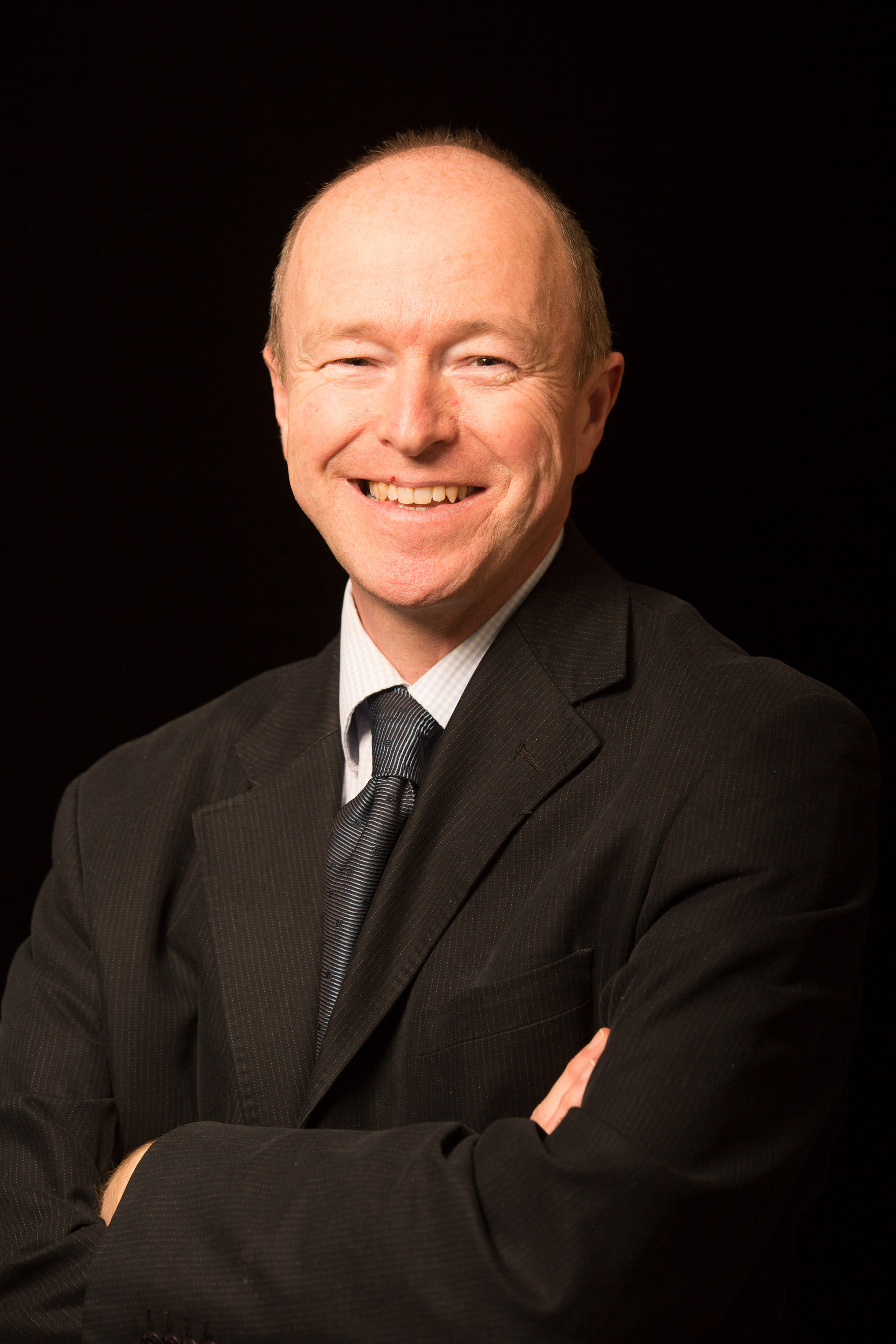}}]{Cormac J. Sreenan} is Professor of Computer Science at University College Cork in Ireland where he leads a research team working on wireless and mobile networking, funded by Science Foundation Ireland, the EU, Enterprise Ireland, and numerous companies. Prior to joining UCC he was a research scientist at AT\&T (Bell) Labs in NJ USA. He has a Ph.D. from the University of Cambridge, and is a Fellow of the Irish Academy of Engineering, a Life Fellow of the Cambridge Philosophical Society, and a Senior Member of the IEEE.
\end{IEEEbiography}
\vskip 0pt plus -1fil 
\begin{IEEEbiography}[{\includegraphics[width=1in,height=1.75in,clip,keepaspectratio]{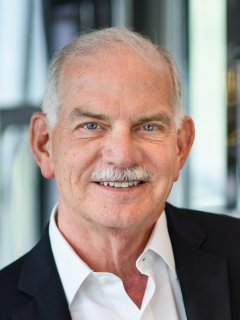}}]{H. Vincent Poor} (S’72, M’77, SM’82, F’87) received the Ph.D. degree in EECS from Princeton University in 1977.  From 1977 until 1990, he was on the faculty of the University of Illinois at Urbana-Champaign. Since 1990 he has been on the faculty at Princeton, where he is currently the Michael Henry Strater University Professor. During 2006 to 2016, he served as the dean of Princeton’s School of Engineering and Applied Science, and he has also held visiting appointments at several other universities, including most recently at Berkeley and Caltech. His research interests are in the areas of information theory, machine learning and network science, and their applications in wireless networks, energy systems and related fields. Among his publications in these areas is the book Machine Learning and Wireless Communications. (Cambridge University Press, 2022). Dr. Poor is a member of the National Academy of Engineering and the National Academy of Sciences and is a foreign member of the Royal Society and other national and international academies. He received the IEEE Alexander Graham Bell Medal in 2017.
\end{IEEEbiography}
\vskip 0pt plus -1fil 
\begin{IEEEbiography}[{\includegraphics[width=1in,height=1.75in, clip, keepaspectratio]{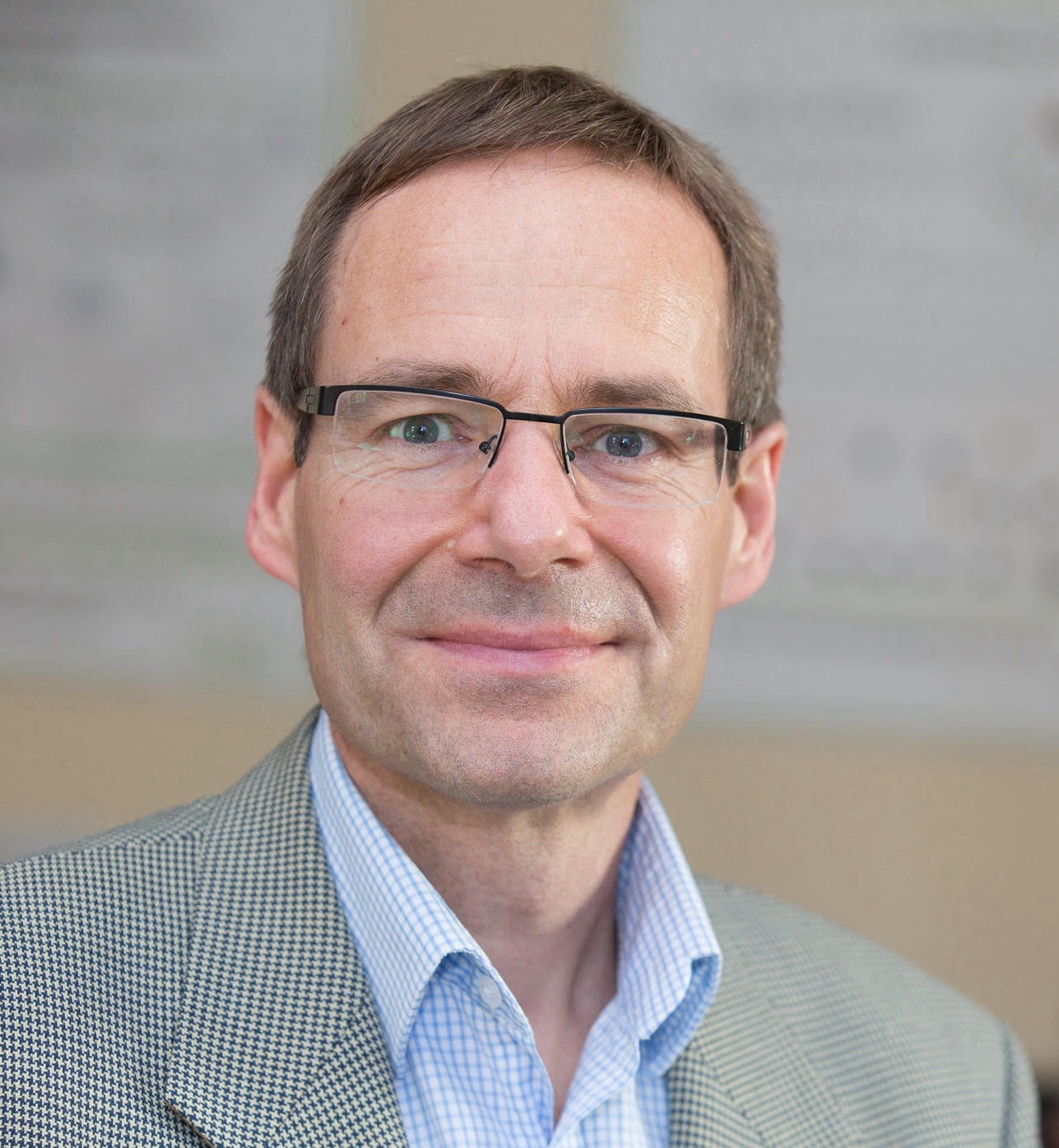}}]{Dirk Pesch}(Senior Member, IEEE) received the Dipl.Ing. degree from RWTH Aachen University, Aachen, Germany, and the Ph.D. degree from the University of Strathclyde, Glasgow, U.K. He is currently a Professor and Head of School of the School of Computer Science and Information Technology at University College Cork, Ireland. Dirk has more than 25 years research and development experience in both industry and academia and has authored or coauthored more than 200 scientific articles and book chapters. His research interests include architecture, design, algorithms, and performance evaluation of low power, dense and mobile wireless networks and services for Internet of Things and Cyber-Physical System's applications in smart connected communities, health and wellbeing, and smart manufacturing. He is a Principal Investigator of the Science Foundation Ireland funded CONNECT Centre for Future Networks and the Director of the SFI Centre for Research Training in Advanced Networks for Sustainable Societies. Dirk contributes to international conference organization and is a Member of the Editorial Board of a number of journals.
\end{IEEEbiography}

\end{document}